\documentclass[lettersize,journal]{IEEEtran}
\usepackage{amsmath,amsfonts}
\usepackage{algorithmic}
\usepackage{array}
\usepackage{cite}
\usepackage[caption=false,font=normalsize,labelfont=sf,textfont=sf]{subfig}
\usepackage{textcomp}
\usepackage{stfloats}
\usepackage{url}
\usepackage{verbatim}
\usepackage{graphicx}
\usepackage{adjustbox}
\def\BibTeX{{\rm B\kern-.05em{\sc i\kern-.025em b}\kern-.08em
    T\kern-.1667em\lower.7ex\hbox{E}\kern-.125emX}}
\usepackage{balance}
\usepackage{xr}
\usepackage{multirow}

\usepackage{url,lineno,microtype,subcaption}
\usepackage{siunitx}
\usepackage{upgreek}
\usepackage{seqsplit}
\usepackage{booktabs}

\usepackage[draft]{hyperref} 
\usepackage[noabbrev,capitalize,nameinlink]{cleveref}

\title{Neural Signatures Within and Between Chess Puzzle Solving and Standard Cognitive Tasks for Brain-Computer Interfaces: A Low-Cost Electroencephalography Study}
\author{Matthew Russell*, Samuel Youkeles, William Xia, Kenny Zheng,
Aman Shah, and Robert J.K. Jacob 
\\Human-Computer Interaction Lab, Tufts University
\\Medford, MA, USA
\\*Corresponding author: mrussell@cs.tufts.edu}

\begin{document}

\maketitle

\noindent © 2025 IEEE. Personal use of this material is permitted. Permission
from IEEE must be obtained for all other uses, in any current or future
media, including reprinting/republishing this material for advertising or
promotional purposes, creating new collective works, for resale or
redistribution to servers or lists, or reuse of any copyrighted
component of this work in other works.\\

\begin{abstract}
Consumer-grade electroencephalography (EEG) devices show promise for Brain-Computer Interface (BCI) applications, but their efficacy in detecting subtle cognitive states remains understudied. We developed a comprehensive study paradigm which incorporates a combination of established cognitive tasks (N-Back, Stroop, and Mental Rotation) and adds a novel ecological Chess puzzles task. We tested our paradigm with the MUSE 2, a low-cost consumer-grade EEG device. Using linear mixed-effects modeling we demonstrate successful distinctions of within-task workload levels and cross-task cognitive states based on the spectral power data derived from the MUSE 2 device. With machine learning we further show reliable predictive power to differentiate between workload levels in the N-Back task, and also achieve effective cross-task classification. These findings demonstrate that consumer-grade EEG devices like the MUSE 2 can be used to effectively differentiate between various levels of cognitive workload as well as among more nuanced task-based cognitive states, and that these tools can be leveraged for real-time adaptive BCI applications in practical settings.
\end{abstract}

\section{Introduction}
Electroencephalography (EEG) has emerged as a powerful tool in Brain-Computer Interface (BCI) research. In particular, EEG has an established capability to measure cognitive workload \cite{Gevins03}, which has proven valuable in ergonomics and human-computer interaction studies for differentiating task difficulties \cite{Zarjam15}. And recent advances in consumer-grade devices have expanded EEG's accessibility beyond traditional clinical settings. This democratization has enabled researchers across diverse fields, from engineering to cognitive neuroscience, to leverage EEG technology in novel ways \cite{sabio_williams_mcarthur_badcock_2024}. 

However, bridging the gap between clinical neuroscience findings, initial validation of low-cost devices, and practical applications for general users remains an active area of research \cite{Varbu22}. Further, while EEG technology continues to advance, significant work is still needed to establish reliable real-time signal processing methods suitable for BCI applications \cite{McFarland17}. These challenges necessitate empirical studies to both validate the technological developments of lower-cost devices, and to inform the design of future user interfaces.

Our research employs the MUSE 2, a wireless and portable EEG device initially validated for cognitive workload measurement in BCI applications \cite{PeiningTan17, Bird18}. Despite this initial validation, there remains limited understanding of its practical application in realistic BCI scenarios. To address this gap, we investigate the device's ability to detect mental workload across four distinct tasks: three well-validated cognitive psychology paradigms (Mental Rotation, Stroop, and N-Back) and a practical Chess\footnote{We will hereafter capitalize ``Chess'' to refer to the task paradigm of playing chess puzzles.} paradigm using chess puzzles. The selection of these four cognitive tasks: N-Back, Stroop, Mental Rotation, and Chess—creates a comprehensive framework for examining distinct yet interrelated aspects of cognitive function: the N-Back task primarily assesses working memory and information manipulation, the Stroop test measures inhibitory control and attention, the Mental Rotation task evaluates spatial reasoning, and the Chess task introduces a unique dimension of strategic thinking and pattern recognition that bridges and extends these cognitive domains. The Chess puzzle task presents a particularly valuable application in this battery because they require the simultaneous engagement of multiple cognitive processes: working memory (similar to N-Back, but in a more complex context), inhibitory control (as in Stroop, but applied to move selection), and spatial reasoning (complementing Mental Rotation, but with added strategic complexity).

Our study provides four key contributions. First, we introduce the Chess task as a new paradigm alongside the other cognitive psychology tasks. Secondly, we benchmark our ability to differentiate between difficulty levels within each task. Third, we examine the differences in neural signatures between tasks, independent of difficulty level. Our analysis explores both aspects through statistical analyses with Linear Mixed-Effects Regression, as well as through classification by way of machine learning to predict on unseen samples. And fourth, we provide the experimental paradigm code and data collected as open-access for future researchers. 

Our work addresses three primary research questions, all in relation to the MUSE 2 device:
\begin{enumerate}
\item Can we reliably detect and classify different levels of cognitive workload within individual tasks (N-Back, Stroop, Mental Rotation, and Chess)?
\item To what extent can we distinguish between different cognitive tasks, independent of their difficulty levels?
\item Are the neural signatures identified in questions 1 and 2 robust enough to support real-time classification for BCI applications?
\end{enumerate}

%
%

\section{Background and related work}

\subsection{Low-Cost EEG for Mental Workload and State Classification}
In recent years, advancements in consumer-grade, low-cost electroencephalography (EEG) devices—such as Muse II, OpenBCI, NeuroSky, InteraXon, and Emotiv—have positioned them as viable alternatives to traditional, expensive EEG systems \cite{vos2024stress}. These devices offer an accessible means of monitoring brain activity, enabling applications in dynamic, real-world settings. Their portability, affordability, and ease of use have made them practical tools for investigating mental workload and cognitive state classification, with demonstrated applications in attention monitoring, stress detection, and adaptive human-computer interaction \cite{knoll2011measuring,bashivan2016mental}.

Building upon this progress and aligning with our related work, studies such as \cite{so2017evaluation} have investigated the potential of single-channel EEG systems, such as NeuroSky, to differentiate nuanced cognitive tasks. Their evaluation of a single-probe frontal EEG system demonstrated its effectiveness in distinguishing tasks such as arithmetic operations, finger tapping, mental rotation, and lexical decisions. Notably, they observed an increase in $\theta$ band power during these tasks and successfully applied support vector machines (SVMs) to frequency-transformed 2.5-second windows, achieving classification accuracies exceeding 70\% for arithmetic operations, finger tapping, and lexical decision. However, their cross-validation procedure did not account for auto-correlated features in time-related samples, highlighting opportunities for methodological refinement.

Devices like Muse II, which incorporate a four-channel dry electrode design, strike a balance between usability and functionality, making them well-suited for real-time applications. Despite this, these systems often face limitations in spatial resolution and signal fidelity when compared to research-grade devices \cite{lee2024evaluation}. To overcome such challenges, advancements in machine learning techniques—such as SVMs, convolutional neural networks (CNNs), and transformer-based models—have significantly improved classification accuracy and enhanced signal processing capabilities \cite{hassan2024eeg,larocco2020lowcost,10171942}.

As these technologies evolve, devices like Muse II highlight the potential to bridge the gap between affordability and performance. Their demonstrated and proven success in cognitive workload classification within controlled paradigms, such as the N-Back task, underscores their value in adaptive systems across domains such as human-computer interaction. With continued innovations in AI-driven analysis, low-cost EEG systems are poised to expand the frontiers of accessible and scalable BCI research.

\subsection{Muse 2 in Research}
Research has found that the MUSE 2 Headband, a portable and wireless device capable of measuring EEG, is a suitable alternative to traditional EEG methods. For example, to test the headband's utility, a visual oddball paradigm experiment and a reward-learning task experiment were conducted. MUSE was able to observe and quantify the N200 and P300 ERP components in both tasks \cite{krigolson2017choosing}. The headband provides real-time feedback and monitoring of EEG indicators, and was validated for its accuracy in measuring EEG signals \cite{hawley2021technology, krigolson2017choosing}.

Preliminary studies have also been done with the MUSE 2 which demonstrate the potential of low-cost EEG systems to classify mental states based on activity level. \cite{Bird18}, for instance, collected data from five participants across three mental states (relaxing, neutral, and concentrating), analyzing five different brainwave frequencies. Using feature selection techniques and machine learning classifiers including Bayesian Networks, Support Vector Machines (SVM) and Random Forests, they achieved over 87\% accuracy in state classification. Another study done by \cite{PeiningTan17} investigated the development of a BCI system to enable drone control through consumer-grade EEG headsets. They focused on adapting drone thrust levels based on concentration, and likewise leveraged SVM and achieved 70\% accuracy, however they also found wide variability among participants' individual scores. In another study, \cite{zhang_muse} tested the reliability of the MUSE 2 device against the TOBII Pro Nano eye tracking system. Their work suggests that the eye tracking system performs better, for classification was of N-Back levels 1 vs. 2. The relatively high cost of the TOBII device, however, likely puts it out of league with the MUSE 2. 

While these studies demonstrate promising advances in consumer-grade EEG classification, they primarily focus on basic state differentiation rather than the complex cognitive processes involved in ecological task scenarios. Our work extends this foundation by examining more nuanced cognitive states across tasks of varying complexity and real-world applicability with low-cost EEG. And, given that one of the objectives of our study is to be as portable, cost-efficient, and accurate as possible, we decided to use the MUSE 2.

\subsection{Tasks}
\subsubsection{N-Back}
The conceptualization of the N-Back task traces its roots to the work of Kirchner in 1958 \cite{Kirchner1958AgeDifferences}, establishing a cornerstone cognitive exemplar for the probing of working memory and executive function. This task is paramount in the realm of cognitive investigation due to its unique capacity to measure the retention and manipulation of information across brief intervals. Through the employment of diverse stimuli—ranging from auditory-verbal to visuospatial—and the increasing complexity introduced by 1-back, 2-back, and 3-back variations, the N-Back task presents an intricate measure of working memory capacity, executive function, and attentional control aptitudes. The operational framework of this task is meticulously designed to assess accuracy, response latencies, and error rates, thereby illuminating the rapidity and proficiency of cognitive operations by requiring participants to simultaneously store and manipulate information \cite{Jaeggi2010ConcurrentValidity, Conway2005, KaneEngle2002}.

\subsubsection{Stroop}
As a widely used practice in psychology, the Stroop Test is a task designed to test the measurement of a user’s cognitive load \cite{gwizdka2010using}. Created by John Ridley Stroop in 1935, the task focuses on specific cognitive processes such as selective attention, memory, learning, cognitive load, and processing speed \cite{macleod1991john}. The test requires the participant to identify the color of a printed word while ignoring the actual word itself. For instance, if the word `BLUE' is printed in a yellow font, the correct response would be `yellow.' The task measures the user’s accuracy and speed to give insight into the user’s cognitive ability and efficiency \cite{macleod2015stroop}. 

\subsubsection{Mental Rotation}
The Mental Rotation Task (MRT) is a cognitive psychological test devised by Roger Shepard and Jacqueline Metzler in 1971, aimed at assessing spatial reasoning ability. In this test, participants are presented with two 3-dimensional objects in different perspectives positioned in space. The objective is to determine whether the objects are identical. Studies utilizing the MRT have revealed that reaction time tends to increase linearly as the angular disparity between the orientations grows, which indicates that participants mentally rotate one of the objects to determine if it matches the other \cite{shepard1971mental}. Moreover, EEG work has demonstrated that the MRT elicits activation in the superior parietal lobule and the intraparietal sulcus, indicating their involvement in spatial processing during the task
\cite{osuagwu2014similarities}. 

\subsubsection{Chess}
Several studies have explored the effects of chess playing on the brain. Increased theta power in posterior brain regions was observed when chess players engaged in faster-paced games, suggesting a link to long-term memory retrieval and chunk processing \cite{santos2019electroencephalographic}. High-level chess players have also been shown to exhibit increased alpha EEG power in the occipital area during chess playing, indicating a greater adaptive response to the cognitive demands of the task \cite{santos2021neurophysiological}. Another study found that winning players exhibited higher theta power in frontal, central, and posterior brain regions as the difficulty of the opponent gradually increased;
higher alpha power was observed in more challenging games, suggesting the engagement of creative thinking and the exploration of alternative solutions. By contrast, losing players demonstrated a decrease in beta and alpha power as the opponent's difficulty escalated, indicating a lack of adaptive response to challenging situations and an inability to formulate effective solutions \cite{fuentes2019chess}. Although one study explored ERP responses to the playing of Chess puzzles \cite{Hageman2021}, the application of a puzzle-paradigm in the context of EEG is nevertheless underexplored. 

\section{Materials and Methods}

\subsection{Study Outline}
We performed our study in two phases
\begin{enumerate}
    \item In our initial phase I we ran participants \textit{either} on the set of mental workload tasks \textit{or} on the Chess puzzle task. These data are leveraged for within-task workload analyses only. 
    \item In phase II we ran a second set of participants \textit{both} on the mental workload tasks \textit{and} the Chess puzzle task in the same sitting. These data are both added to the within-subjects workload dataset from phase I, and also leveraged separately for cross-task analyses. 
\end{enumerate}

The tasks used for both phases I and II were the same. For each participant in both studies we administered a preliminary consent form, fit them with the MUSE device, had them perform the tasks, and informally debriefed them. For phase I, participants for the Chess study were sent a \$10 Amazon gift card, and participants for the mental workload study were sent a \$20 Amazon gift card. For phase II, participants were sent a \$30 Amazon gift card. No participants from phase I were enlisted for phase II. The study was approved by our organization's Institutional Review Board. 

While we use the data from both phases I and II to model within-task difficulty levels, we only use the data from phase II for our cross-task analyses. 

\subsection{Program Implementation}
We implemented our experiments using JavaScript, and leveraged several key libraries and tools.
We built the core experimental framework with \texttt{jspsych} \cite{de2023jspsych}, which provided a flexible foundation for our study design. For the chess-related components, we employed \texttt{chess.js} \cite{chessjs} to handle game logic and \texttt{chessBoard.js} \cite{chessboardjs} to create an interactive visual interface. We interfaced our code with the MUSE 2 device by way of the \texttt{muse-js} \cite{musejs} library. 


\subsection{Chess Puzzles Database}
\label{section:chess_database}
To create a database of Chess puzzles we began by downloading the \texttt{lichess.org} open chess puzzle database, which contains millions of Chess puzzles \cite{lichess2024}. To minimize confounding factors due to puzzle differences, we filtered the puzzles to only contain those that result in checkmate, which resulted in a total of 686,559 puzzles. Each puzzle had an associated Glicko2 rating to quantify its difficulty \cite{glickman2012example}. We further filtered the total puzzle set based on Glicko2 difficulty ratings within the range [600, 2250], where across the \texttt{lichess.org} site a rating of 600 represents the skill level of a complete beginner and the rating of 2250 represents the skill level of an expert \cite{lichessWeeklyClassical}. We then binned the puzzles into Glicko2 rating ranges of 50; the first bin contained puzzles with ratings from [600-650), and so on. The lower rating bins (600-1750) each contained over 10,000 puzzles. While the higher rating bins (1800-2250) contained fewer puzzles, each bin still maintained a minimum of 197 puzzles, ensuring sufficient availability for high-performing participants. 

\subsection{Task Details}

\subsubsection{Chess Task}
\begin{figure}[h!]
\centering
\includegraphics[width=\linewidth]{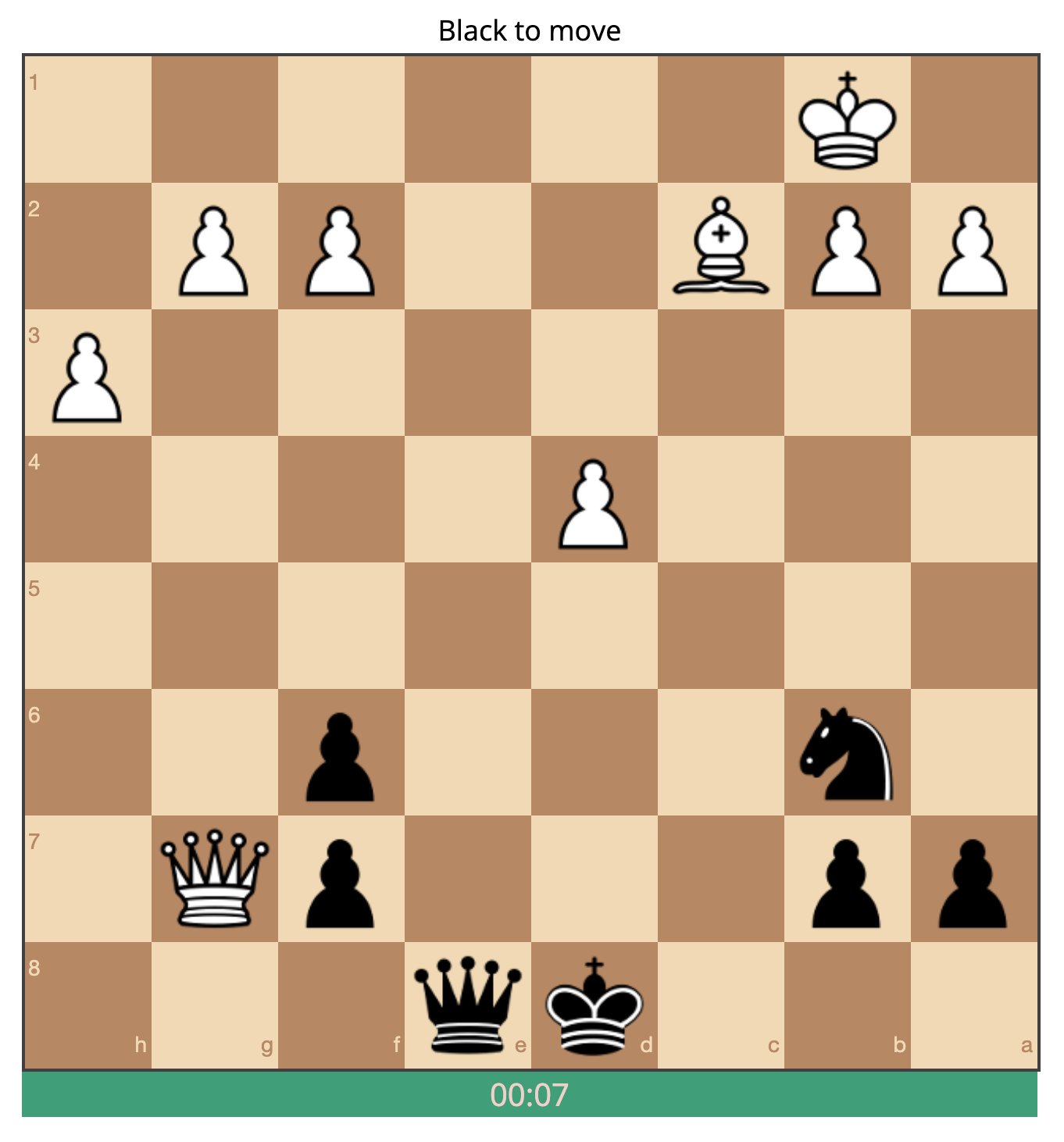}
\caption{A chess puzzle where the solution is to move the black queen to square e1, and then to square d1, which delivers checkmate by taking the white bishop that will move to defend the white king.}
\label{fig:CHESS_PUZZLE}
\end{figure}

In the Chess task, participants played a total of 6 rounds with 30 puzzles each. Each puzzle consisted of an arrangement of pieces such that the participant could solve the puzzle by making a certain sequence of moves to reach a checkmate. An example of a puzzle is shown in Figure \ref{fig:CHESS_PUZZLE}. The first puzzle of each round started with a Glicko2 rating from the 800-850 bin. If a participant successfully completed a given puzzle, the next puzzle would come from the next hardest bin. Conversely, if the participant made an incorrect move while solving a puzzle, the next puzzle would come from the next easiest bin. After each move, participants were shown whether their move was correct. Participants were instructed to perform the puzzles as quickly as possible without sacrificing accuracy. The maximum time allotted per-puzzle was 30 seconds.

\subsubsection{Cognitive Neuroscience Tasks}
We ran the three cognitive tasks (N-Back, Stroop, and Mental Rotation) in a random order. Before completing each cognitive task, the participants were provided with a written explanation of the task. Next, they performed a task demo where they practiced the tasks and were shown after each response whether or not their answers were correct. During the training period, participants were encouraged to ask questions to ensure they understood the task; if they were confused about how to complete the task, they had the opportunity to perform the training session again. No participants performed the training session for any task more than 2 times. For each trial across all task types, a fixation cross was displayed for 500ms prior to trial onset, the inter-trial interval was 250ms, and a keyboard response ended the trial.

For the N-Back task, participants were shown a series of letters. Each letter was shown for 500ms, and the participant had 2500ms to respond, after which the trial ended automatically. For a given trial, they were instructed to press the \texttt{y} key if the current letter was the same as the letter they saw N letters previously (known as a N-Back trial), or to press the \texttt{n} key otherwise; N-Back trials occurred with 30\% frequency in all workload conditions. See Figure \ref{fig:NBACK_BLOCK} for an example of a series of N-Back trials. 

\begin{figure}[h!]
\centering
\Huge
B F G T \textcolor{blue}{G} G \textcolor{blue}{G} T X \textcolor{blue}{T} L
\caption{ A series of possible characters presented during a N-Back task. During the study, the characters are shown one at a time. Blue colored letters indicate 2-Back targets. That is, on seeing all of the first four letters, the participant would press the `N' key. However, the 5th character `G' was also seen two characters prior, thus the participant would press `Y'.  }
\label{fig:NBACK_BLOCK}
\end{figure}

For the Stroop task, participants were shown a series of words that each spelled the name of a color – these words were also written in a text color which was not necessarily the same as the color they spelled. The participant pressed the key corresponding to the first letter of the word's text color. Congruent trials, in which the color of the word matches the color that the word spelled, occurred with 50\% frequency. Each stimulus was shown for a maximum of 2000ms, after which the trial ended automatically. See Figure \ref{fig:STROOP_BLOCK} for an example of an incongruent trial.

\begin{figure}[h!]
\centering
\Huge
\textcolor{blue}{red}
\caption{ Example of an incongruent Stroop stimulus; the word `red' is written in blue. In this case the participant would press the `b' key on their keyboard, indicating the color the word is written in, not the color that is spelled. Red, yellow, blue, and green were all used for the study.}
\label{fig:STROOP_BLOCK}
\end{figure}

\begin{figure}[h!]
\centering
\includegraphics[width=\linewidth]{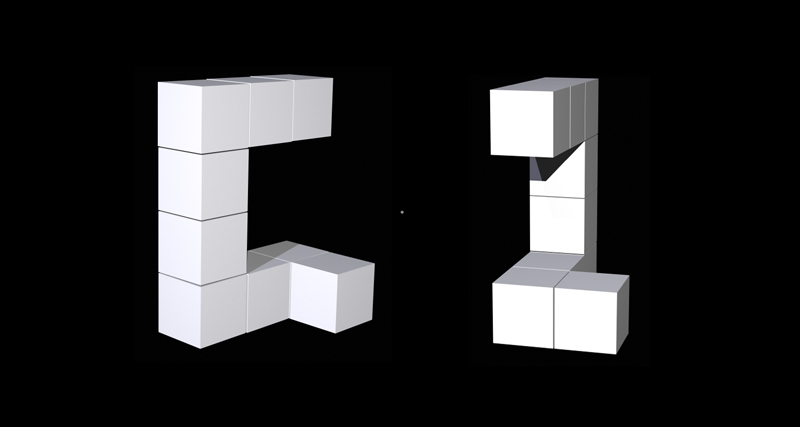}
\caption{ Example of a trial used for the Mental Rotation task (images used from work of \cite{ganis2015new}). This is not a valid rotation, because the second object is not a rotated version of the first object.}
\label{fig:ROTATION_BLOCK}
\end{figure}

For the Mental Rotation task, participants were shown a series of pairs of 3D blocks. We used the block paradigm and images developed by \cite{ganis2015new}. Participants were asked to press the \texttt{y} key if the blocks were the same but rotated, and the \texttt{n} key if they were not the same block. Blocks were the same with 50\% probability. Each stimulus was shown for a maximum of 7500ms, after which the trial ended automatically. See Figure \ref{fig:ROTATION_BLOCK} for an example of the Rotation task. 

\begin{figure}[h!]
    \centering
    \includegraphics[width=.93\linewidth]{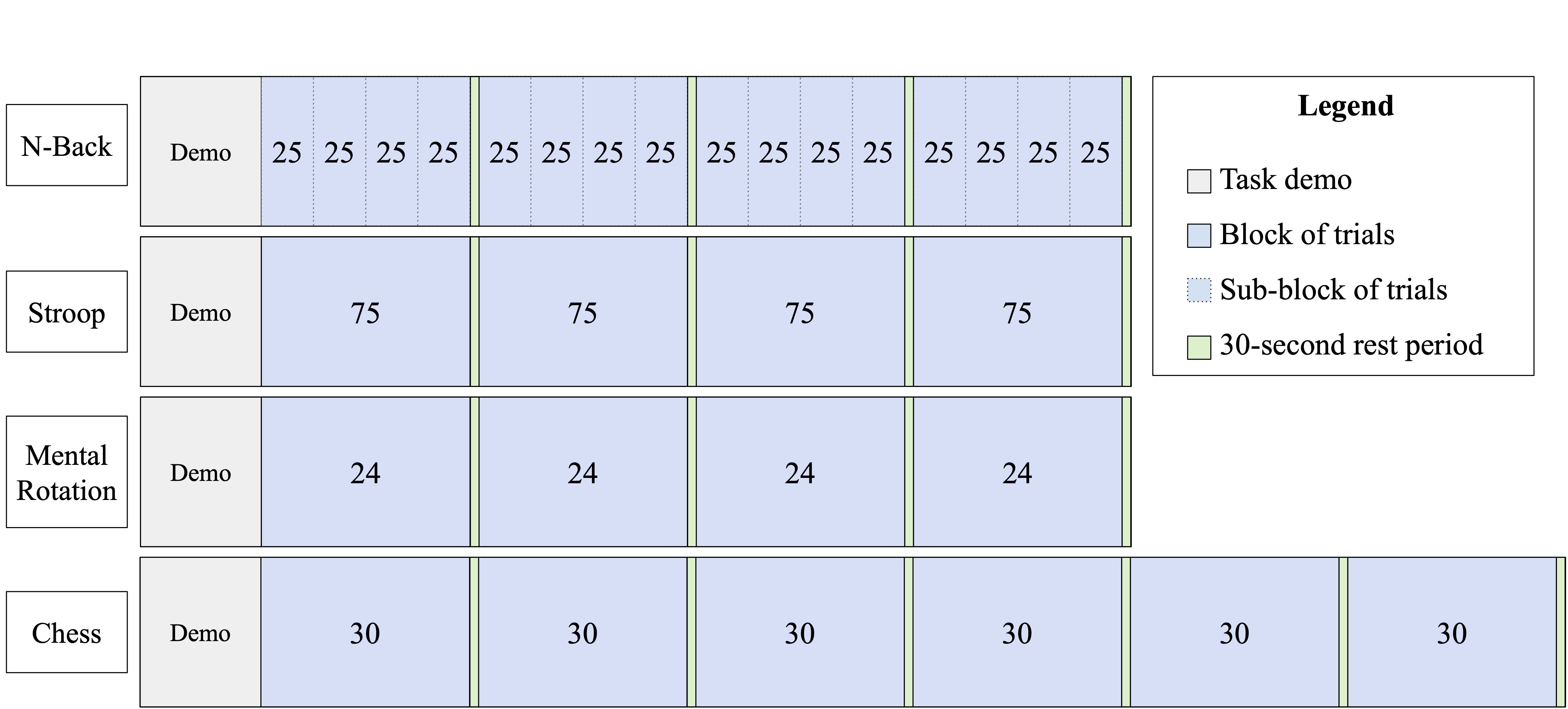}
    \caption{Experimental procedure per task. In the N-Back task, each set of 4 sub-blocks are a random permutation of 25 N-Back trials of the same N, where N is selected from \{0,1,2,3\}. Stroop blocks are 75 trials each, and Mental Rotation blocks are 24 trials each. The Chess blocks are each 30 puzzles in length. Note that this graphic only indicates the number of trials collected; due to data loss from the MUSE 2, time-per-trial, and number of trials per-block, our final dataset resulted in imbalanced samples (see  \cref{supp-tab:DATA_DESCRIPTION} in the supplementary material). }
    \label{fig:EXP_PROCEDURE}
\end{figure}

Four blocks of trials were run for each cognitive task. Each block of the Stroop task contained 75 trials; each block of the Mental Rotation task contained 24 trials; each block of the N-Back task contained four sub-blocks of 25 trials each, one per N-Back variant, in a randomized order. See Figure \ref{fig:EXP_PROCEDURE} for a visual representation of the experimental paradigm. After each block, there was a 30-second rest. Note that the combination of different maximum time per-trial, along with different number of trials-per-block, resulted in different total numbers of trials and epochs-per-trial, collected across participants. 

\subsection{Participant Information and Chess Player Skill}
Our subject population is composed by healthy graduate and undergraduate students from Tufts University (overall ages range from 18-26, mean 20.4). In phase I, our subject pool involved 8 Chess participants (1 female) and 13 mental workload task participants (4 female). For phase II, we recruited 9 participants (all male) to do both tasks in one sitting. Our workload-specific dataset therefore begins with 17 Chess participants and 22 mental workload task participants, and our cross-task analysis dataset contains 9 participants.  

Chess participants also filled out a form indicating the frequency of their Chess play. The distribution of responses is shown in \cref{supp-fig:PLAY_FREQUENCY} in the supplementary material - a substantial majority (75\%) of participants responded that they play at least once per week. After gathering our data, we assessed skill distribution of Chess players by examining the maximum puzzle difficulty each participant successfully completed, as illustrated in \cref{supp-fig:MAX_ELO} in the supplementary material. For reference, puzzles rated at 1400 Elo are the most frequently attempted puzzles in the lichess.org database \cite{lichess2022}. While participant skill level could theoretically influence neural responses during puzzle solving, our preliminary analyses found no significant effects. Given these results, subsequent analyses do not investigate skill-based effects. Future research examining skill-based neural responses may benefit from recruiting a larger participant pool with broader representation across skill levels.

\subsection{EEG Data Collection and Preprocessing}
\begin{figure}[h!]
\centering
\includegraphics[width=.75\linewidth]{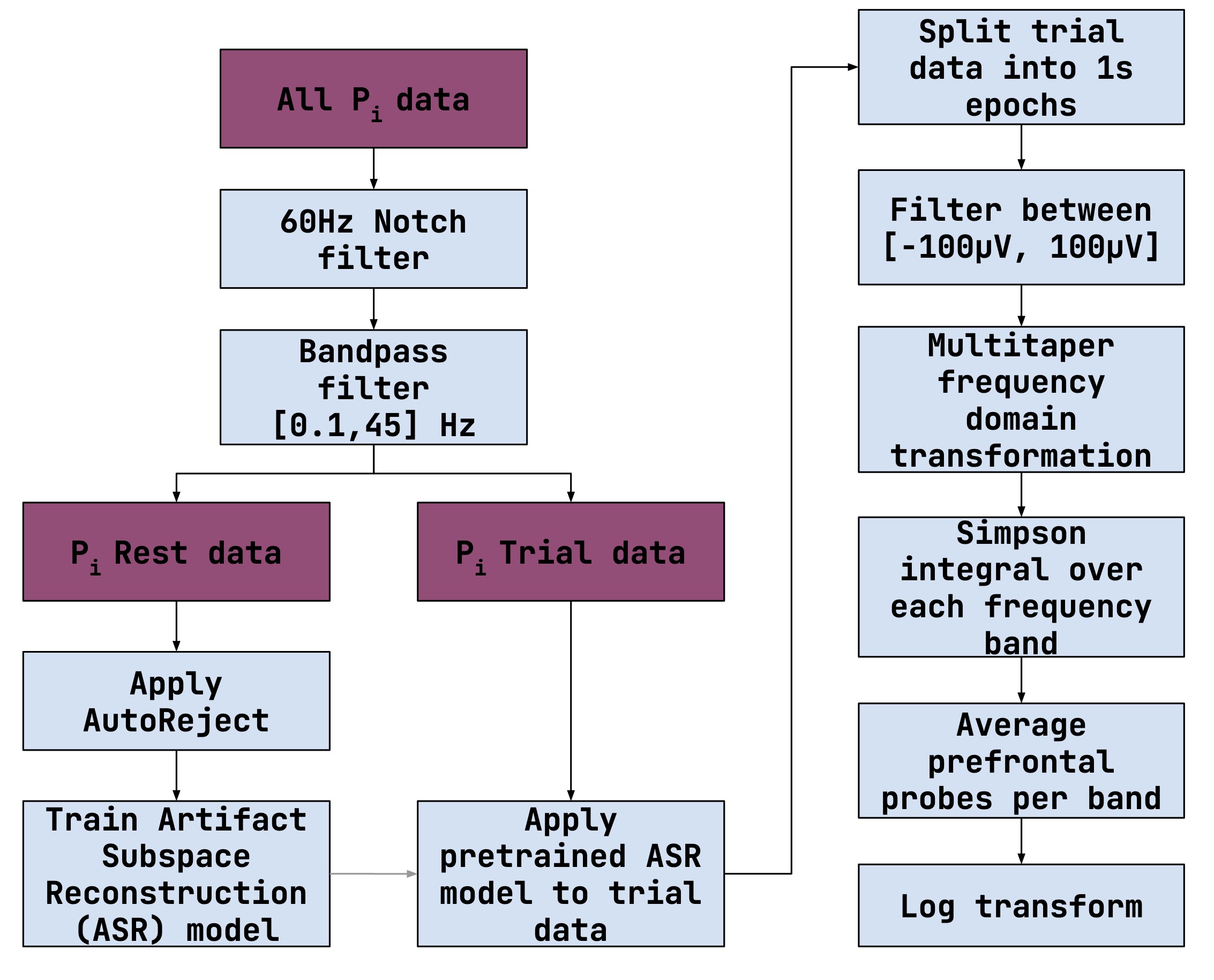}
\caption{Visualization of the EEG processing pipeline showing data flow from raw signal through filtering, artifact removal, and spectral analysis stages for a single participant $P_i$. Rest data is cleaned with the AutoReject algorithm, and the cleaned data is used to train an ASR model, which is then applied to trial data before frequency domain transformation and averaging.
}
\label{fig:PREPROCESSING_PIPELINE}
\end{figure}

Raw data was collected via \texttt{muse-js} at 256 Hz. We focus our analysis only on the prefrontal cortex data, accessed through the MUSE 2's AF7 and AF8 probes. Our primary objective in the EEG analysis was to preserve the maximum amount of usable data while developing a methodology which could be leveraged in the future for online processing. Data pre-processing and feature extraction were both conducted in Python. Data preprocessing are visualized in Figure \ref{fig:PREPROCESSING_PIPELINE}.

We first utilized the \texttt{mne} package \cite{GramfortEtAl2013a} to implement a notch filter at 60 Hz and a bandpass filter spanning 0.1 to 45 Hz\footnote{We initially included frequencies as low as 0.1Hz to accommodate potential ERP analyses \cite{tannerFilter}, but ultimately focused on $\geq$4Hz for our frequency analysis. Consequently, frequencies below 4Hz were not used in our final analytical pipeline.}. For each participant, we applied the AutoReject algorithm \cite{Jas2017Autoreject} on their resting-state data, which we automatically split into epochs of 2 seconds for preprocessing. \footnote{The AutoReject function from the python pip package autoreject (version 0.4.3) was used, with default parameters, and a fixed random state for reproducibility.} AutoReject fully automates the correction of momentary and transient movement artifacts by establishing sensor-specific thresholds based upon effective cross-validation techniques across multiple datasets (the MNE sample data \cite{Gramfort_2013}, the multimodal faces dataset \cite{WakemanandHenson_2015}, and the EEGBCI motor imagery data \cite{Goldberger_2000, Schalk_2004}) to help distinguish genuine neural signals from artifacts caused by muscle movements, eye blinks, or technical issues. Once thresholds are defined, AutoReject evaluates each EEG trial and either repairs by interpolating clean data from adjacent sensors (also known as channels) or excludes them entirely from the dataset to ensure only the cleanest, artifact-free data are to be used in the later stages of analysis. We used the resulting cleaned resting-state data to train an Artifact Subspace Reconstruction (ASR) model \cite{Chang2018ArtifactSubspace, mullen2015real} on their task data\footnote{The ASR function from the python pip package meegkit (version 0.1.7) was used, with default parameters except for the frequency of the input, which was set to 256 to match the MUSE 2 device.}. This model effectively identifies and reconstructs artifact-contaminated subspace segments, thereby ensuring minimal loss of brain signal integrity while preserving a maximum amount of data. The ASR algorithm plays a crucial role by intelligently isolating and reconstructing the data segments contaminated with artifacts, not only enhancing the quality of EEG data by minimizing interference but also preserving the valuable brain signals needed for accurate analysis. We then split the data into epochs (discussed in Section \ref{epoch} below). To optimize signal-to-noise ratio we applied a thresholding procedure \cite{HAPPE} where we excluded any epochs containing values outside the range [-\SI{100}{\micro\volt}, \SI{100}{\micro\volt}] in either of the frontal probes. The median data inclusion rate per-participant is 98\% across both Chess and mental workload datasets, indicating that most participants were largely unaffected by filtering. However, the mean data inclusion rate per-participant is 64\%, indicating that the process disproportionately affected a subset of participants. Participants with excessive data loss ($>=$60\% epochs excluded, n=8) were omitted from the analysis. Of these participants, five were from phase I during the workload tasks, one was excluded from phase I during the Chess tasks, and two were excluded from the phase II data. Although the specific causes of these grouped losses cannot be definitively determined from the available data, the clustering of exclusions suggests potential systematic factors in data collection. For future work, we recommend implementing real-time data quality monitoring to identify issues promptly. Our established preprocessing pipeline can also be adapted for real-time analysis to detect batches of invalid samples as they occur, enabling immediate adjustments to headband positioning or facilitating the collection of additional resting-state data for cleaning procedures.

\subsubsection{Epoching the Data}
\label{epoch}
The selection of epoch length presents a methodological trade-off between signal quality and practical utility. While longer epochs enhance signal-to-noise ratio and provide more reliable spectral estimates, they pose implementation challenges for real-time neural interfaces and provide less information for machine learning models to understand underlying patterns. Additionally, the temporal structure of our experimental paradigm constrains the maximum feasible epoch length: the Stroop, Rotation, and Chess tasks contained multiple workload conditions within single experimental blocks. To optimize for potential real-time classification applications while maintaining adequate signal quality, we extracted epochs from the continuous data stream using fixed time windows. Each epoch was labeled according to the concurrent workload condition. 
We selected 1-second epochs to maximize the number of available samples for machine learning while maintaining sufficient duration for reliable spectral estimation of frequencies as low as 4 Hz. 




\subsubsection{Frequency Domain Transformation and Prefrontal Averaging}
Each epoch was transformed into the frequency domain with \texttt{mne}'s Multitaper library \cite{percival1993spectral, candy2019multitaper}. Following the work of \cite{so2017evaluation}, we delineated frequency bands as follows: $\theta$ (4, 8), $\alpha_1$ (8, 11), $\alpha_2$ (11, 14), $\beta_1$ (14, 25), $\beta_2$ (25, 35), $\gamma_1$ (35, 40), and $\gamma_2$ (40, 45)\footnote{ While we recognize that these are not necessarily the most common waveband delineations used, they represent a similar paradigm within this work done in \cite{so2017evaluation}.}. The Simpson integral \cite{fdez2021cross} was used to extract single values representing the total power in each frequency band. Given our focus on overall pre-frontal activation patterns, we averaged the frequency-domain data from both probes into a single value per frequency bin, optimizing signal-to-noise ratio while maintaining measurement validity for our research objectives. Resulting power values were log-transformed. 


\subsubsection{Data Collection Challenges and Solutions}
\label{exclusion}
Throughout our study we experienced connectivity issues with the MUSE device. In our initial implementation of the experimental protocol, this would cause complete data loss for a participant. To mitigate data loss, we implemented a reconnection protocol prior to the start of each task block (e.g., one block of N-Back, Stroop, mental workload, or Chess puzzles). This approach significantly reduced data loss overall, but some blocks were still lost for some participants. \cref{supp-tab:DATA_DESCRIPTION} in the supplementary material presents detailed information on the number of captured trials and epochs captured per-participant per-task after data loss was taken into account. 


For future research using the MUSE 2 device, we propose several methodological refinements to enhance data quality and reliability. First, we strongly recommend utilizing a wired connection for data transmission rather than wireless connectivity. While this may introduce some constraints for ecological validity in real-time applications, our experience indicates that data loss represents a critical limitation that must be addressed to ensure the proper collection of data. Second, we advocate for the implementation of systematic signal quality assessment protocols. These should include initial verification of proper headband positioning and continuous monitoring of signal integrity throughout the experimental session as discussed above. Such proactive quality control measures would allow researchers to identify and address technical issues promptly, thereby minimizing data loss and reducing reliance on post-hoc exclusion criteria. It is important to note that the current iteration of the \texttt{muse-js} acquisition framework lacks native support for automated signal quality assessment. Consequently, researchers may need to develop custom solutions to implement these recommended quality control measures. Despite this limitation, we believe these procedural modifications will substantially improve data integrity in future studies utilizing this device.

\subsection{Workload Labels}
For our initial statistical analyses we observe the distinctions among reaction time and correctness for the tasks with multi-level classification where possible: for N-Back, difficulty is the level of the N-Back (0-back, 1-back, 2-back, or 3-back). For Rotation, difficulty is the degree of rotation (0 degrees, 50 degrees, 100 degrees, or 150 degrees). For Stroop, difficulty is congruent (0) or incongruent (1), and for Chess, difficulty of a given puzzle for a given participant is the quartile expression (labeled as 0-3) of that puzzle’s difficulty rating within all of the unique puzzle difficulty ratings encountered by that participant. For the machine learning analyses we simplify our labels by grouping all workload samples within a trial type in [0, 1] workload range as low workload (labeled 0), and grouping all workload samples within a trial type in [2, 3] workload range as high workload (labeled 1). Although in this orientation of the data we run the risk of classifying samples distinctly from categorizations of workload classification which are rather close together, we believe that, particularly given the data loss suffered from the MUSE 2 device, the larger sample size per group is a better choice for this analysis. We encourage others to apply their own techniques to the data, which will be made publicly available.

\subsection{Statistical Methodology}
\label{section:statmethods}
The initial part of our analysis employs statistical methods to examine the measured neurophysiological data for both within-task workload and between-tasks. Except where otherwise stated, we utilize Linear Mixed-Effects models to achieve this aim. In each model, we incorporate Participant ID and Task Block per-participant as grouping factors for random effects. We implemented the models with R's \texttt{lmerTest} package \cite{lmerTest}, and validated them through diagnostic checks that included verification of approximate normality of residuals with Quantile-Quantile (Q-Q) plots, posterior predictive checks for model fit, and examination of homoscedasticity and linearity against fitted values using the \texttt{performance} package \cite{performance}. To quantify the magnitude of the fixed effects, we calculate partial eta-squared ($\eta_{p}^{2}$) values using the \texttt{effectsize} library \cite{effectsize}. Estimated marginal means and post hoc pairwise comparisons were conducted with the \texttt{emmeans} library \cite{emmeans}. We used the conventional threshold of $\alpha = 0.05$ for all statistical tests in our analysis.

\subsection{Modeling of Reaction Time and Correctness}
To investigate behavioral metrics we modeled reaction time as a function of difficulty for all of the trials over all participants. For the N-Back, Stroop, and Rotation tasks, time is defined as the reaction time taken to press a key following the stimulus presentation. For the Chess puzzle task, we define time as the time required to solve the puzzle. Reaction time data was log-transformed for statistical modeling \cite{lo2015transform}. Omnibus test p-values for per-task models were corrected with Benjamini-Hochberg correction, as were all post-hoc comparisons. Prior to correction, pairwise comparisons were Tukey-adjusted to control the family-wise error rate. We also modeled correctness as a function of difficulty. Because correctness of a given trial is binary, we employed a Generalized Linear Mixed-Effect Model (\texttt{glmer}) with a binomial family function for the analysis; random effects structure is the same as described in Section \ref{section:statmethods}; p-values were adjusted using the same method as for the Reaction Time tests. Data visualizations are created using the \texttt{seaborn} library \cite{Waskom2021}, and display group means accompanied by their 95\% confidence intervals, which were calculated using bootstrapping methods that account for the hierarchical structure of the data.

\subsection{Statistical Modeling of EEG Data}
Prior to beginning our EEG analysis, we first averaged each participant's data for each unique combination of task type, block, and workload level. The decision to apply this averaging was due to two primary concerns: first, it addresses potential autocorrelation issues arising from the temporal dependencies inherent in time-series EEG data. Second, it enhances the signal-to-noise ratio, which is particularly important when working with consumer-grade EEG devices that typically exhibit higher baseline noise levels than research-grade systems. To analyze workload data within each task, data for each combination of participant, trial type, and wavelength were scaled to zero mean and unit variance. We then created individual models for each unique combination of wavelength and task type. Benjamini-Hochberg correction is used for the omnibus tests across wavelengths within each task type. For each significant model, pairwise comparisons were generated, correcting within wavelength-task model with Tukey adjustment. Benjamini-Hochberg correction is likewise used to correct post-hoc comparisons within a given task type. To analyze EEG data across tasks, individual models were created for each wavelength. For these models, workload was ignored. Benjamini-Hochberg correction is used both across all omnibus tests. For each significant model, post-hoc tests are run and initially corrected with Tukey adjustment, followed by further Benjamini-Hochberg correction across all such post-hoc tests. 

\subsection{Machine Learning Analyses}
To simulate real-world application in an applied BCI context, we also did machine learning analyses on the data. For the machine learning analyses, we do not average the data, but instead use all the available 1-second frequency-domain samples.  This is with an eye towards future applications involving real-time classification. Further, to focus our analysis, we concentrate on the fundamental binary classification problem as a first step in developing reliable real-time BCI applications. Specifically, we relabeled the data from the N-Back, Chess, and Rotation tasks such that the lower two classification levels are considered ``Low" workload, and the higher two classification levels are considered ``High" workload. There are seven features, one for each of the log-transformed power band spectra averaged between the two frontal probes. As with our statistical modeling, we approach the machine learning analysis in two contexts: within-task workload, and cross-task classification. 

Our machine learning modeling works under the assumption that a real-time BCI application will be developed with some training data collected from each individual - that is, rather than attempt leave-one-out cross validation, we allow some data from each participant to be put into the training set. Crucially, however, in order to prevent the models from learning auto-correlated features within each time block, we create our training and testing sets based on blocks of trials as shown in Figure \ref{fig:EXP_PROCEDURE}. That is, any given block of trials taken together will only be either in the test or training set. Notably, each block contains approximately equal samples for each workload class.

\subsubsection{Grouped and Subsampled Monte Carlo Cross-Validation}
Due to data loss from the MUSE device, the fact that the Chess task was collected for as long as the other three tasks combined, and due to varying time-spans of the other three tasks, our dataset contains imbalanced samples both within and across participants. However, this imbalance is not representative of the brain data itself; it stems solely from the experimental paradigm. To develop an online algorithm for differentiating between brain-states or workload levels in practice, we would modify the protocol to create a balanced dataset. Therefore, to test our data with machine learning in the fairest way possible we employed a grouped and subsampled Monte Carlo Cross-Validation methodology for both the within-task workload machine learning models and for the between-task model. The data preparation pipelines are shown in detail in the supplementary material in \cref{supp-fig:ML_PIPELINE_WITHIN_TASK_WORKLOAD} and \cref{supp-fig:ML_PIPELINE_CROSS_TASK}. This method involves 1,000 iterations for the cross-task workload model, and 1,000\footnote{For all of our models, after conducting 1,000 iterations of Monte Carlo Cross-Validation we determined that the 95 confidence interval of the mean fell within ±0.5 percentage point.}, where each iteration follows these steps: 
\begin{enumerate}
\label{data_split}
    \item For each participant
        \begin{enumerate}
            \item Split their data into train and test sets, where 20\% of blocks as described in Figure \ref{fig:EXP_PROCEDURE} are used for testing. This grouping is crucial to avoid the model learning auto-correlated feature patterns from within a given trial block. 
            \item For both test and train sets: subsample all but the minority classes randomly and without replacement. For within-class workload, separate models are produced for each trial type, so a participant's data is subsampled based on the number of samples for each workload level; for cross-task classification, the data from all trial types for a given participant are subsampled based on the number of samples for each trial type.
            \item Scale the data by calculating $\mu$ and $\sigma$ for each feature in the training set; subtract $\mu$ and divide by $\sigma$ for both training and testing data. 
        \end{enumerate}
    \item Balance the number of training and testing samples by subsampling all of the training and testing data (separately and without replacement) based on the combination of participant id and class label. 
    \item Initialize a Random Forest classifier using the \texttt{ensemble.RandomForest} function from \texttt{scikit-learn} \cite{scikit-learn} on the training data using default parameters \footnote{Model configuration included 100 trees (n$\_$estimators), unlimited maximum depth with nodes expanded until reaching pure leaf or minimum split threshold, minimum samples for split of 2, minimum samples per leaf of 1, and Gini criterion for measuring split quality. We opted for default parameters to establish a robust baseline implementation while avoiding potential overfitting from extensive parameter tuning}; record both overall results as well as results of per-participant classification. 
\end{enumerate}

\subsubsection{Within-Task Workload Machine Learning}
For within-task workload we model each task separately. Because in this case we do not need to classify across tasks, we can leverage the entire dataset, including all valid data from participants in both phases I and II. See \cref{supp-tab:PPM_WORKLOAD} in the supplementary material for detailed information on average train and test set sizes across the Monte Carlo iterations. 

\subsubsection{Cross-Task Machine Learning}
For cross-task analysis we require data for each participant from all tasks, therefore we use only the data collected in phase II for this machine learning set. For this machine learning set, we had a total of 6 participants. Across all Monte Carlo iterations we had an average of 2,723 training samples and 776 testing samples. Due to the data splitting process discussed in \ref{data_split}, the training and testing samples were each evenly balanced within each participant for samples per label, and were likewise balanced for samples per-participant. \texttt{scikit-learn} \cite{scikit-learn} was used to implement the grouping strategy, Random Forest model, and correctness statistics. \texttt{imblearn} \cite{imblearn} was used for subsampling of the data. We report the averages over all 1,000 models of the precision, recall, and macro F1-score.

\section{Results}

\subsection{Reaction Time and Correctness}

See \cref{supp-fig:BEHAVIORAL_STATS} for a visualization of reaction time and correctness data. Regarding temporal statistics, for all tasks, time to solve was significant: N-Back ($F_{3, 6166.49} = 795.91$, $p < 0.001$, $\eta^2_p = 0.28$)\footnote{All p-values reported inline reflect adjustments for multiple comparisons. Tables present both unadjusted (p) and adjusted (p.adj) values}, Chess ($F_{3, 2906.35} = 300.37$, $p < 0.001$, $\eta^2_p = 0.24$), Rotation ($F_{3, 1554.14} = 127.65$, $p < 0.001$, $\eta^2_p = 0.20$), and Stroop ($F_{1, 4894.41} = 521.31$, $p < 0.001$, $\eta^2_p = 0.08$). Likewise, with the exception of N-Back 2 vs. 3 (t$_{6114.86}$=-2.18, p=0.128, $\eta^2_p$=0.0), the time required to solve each task increased significantly as difficulty increased. See \cref{supp-tab:BEHAVIORAL_TIME_POST_HOC} in the supplementary material for full post-hoc contrast results. In-person informal conversations with many participants indicated that they were feeling extremely challenged by 3-back such that even `giving up' during the task at parts was common; to this end, the finding of similar reaction times during 2-back and 3-back is sensible. 

Correctness was likewise significantly correlated with workload for all tasks  N-Back ($\chi^2(3, $ N=6225$) = 263.89$, $p < 0.001$), Chess ($\chi^2(3, $ N=2938$) = 244.88 $, $p < 0.001$), Rotation ($\chi^2(3, $ N=1608$) = 31.50$, $p < 0.001$), and Stroop ($\chi^2(1, $ N=4950$) = 56.20$, $p < 0.001$). However, correctness did not have as strong of an effect across workload within tasks. See \cref{supp-tab:BEHAVIORAL_CORRECTNESS_POST_HOC} in the supplementary material for post-hoc contrast results. Correctness was significantly different with $p<0.001$ across all contrasts of Chess, for the Stroop contrast, all of N-Back outside of 0-1 ($z=0.97$, $p=0.765$), and for Rotation contrasts 0-3 and 2-3. Rotation contrast of 1-3 was significant ($z=3.09$, $p=0.014$). However, Rotation contrasts of 0-1 ($z=2.12$, $p=0.172$), 0-2 ($z=1.12$, $p=0.754$), and 1-2 
($z=-1.04$, $p=0.765$) were not significant. The results for correctness were as-expected for Chess and Stroop, and largely the case for N-Back. However, given that 0-back and 1-back are effectively trivial tasks, near-perfect accuracy is sensible here. The Rotation task data, however, indicates that the largest degree of rotation (hardest difficulty) was the only one truly distinct from the others in terms of correctness. 

\subsection{Statistical Analysis of Task-Specific Cognitive Load in EEG Signals}

Full overall results are shown in the supplementary material in \cref{supp-tab:WITHIN_EEG_ANOVAS} and \cref{supp-fig:WITHIN_TASK_WORKLOAD}.

\subsubsection{Chess}

We observed significant workload effects across all relatively high frequency bands in the Chess task:  $\beta_1$ ($F_{3,204.91} = 6.27$, $p = 0.001$, $\eta^2_p = 0.08$), $\beta_2$ ($F_{3,204.27} = 13.40$, $p < 0.001$, $\eta^2_p = 0.16$), $\gamma_1$ ($F_{3,205.27} = 10.29$, $p < 0.001$, $\eta^2_p = 0.13$), and $\gamma_2$ ($F_{3,205.98} = 8.35$, $p < 0.001$, $\eta^2_p = 0.11$). Post-hoc analyses (see \cref{supp-tab:WITHIN_EEG_PHOCS_CHESS} in the supplementary material) revealed the most substantial effects in all frequency bands between the largest gaps in workload levels. For the contrast (0-3): $\beta_1$: $t_{197.61}=-3.96$, $p=0.002$, $\eta^2_p = 0.07$; $\beta_2$: $t_{197.78} = -6.01$, $p < 0.001$, $\eta^2_p = 0.15$; $\gamma_1$: $t_{197.80}=-5.12$, $p < 0.001$, $\eta^2_p = 0.12$; $\gamma_2$: $t_{197.82}=-4.49$, $p = 0.001$, $\eta^2_p = 0.09$), and the contrast (0-2): $\beta_1$: $t_{197.01}=-3.01$, $p=0.038$, $\eta^2_p = 0.04$; $\beta_2$: $t_{197.01} = -3.79$, $p = 0.004$, $\eta^2_p = 0.07$; $\gamma_1$: $t_{197.01}=-3.26$, $p = 0.019$, $\eta^2_p = 0.05$; $\gamma_2$: $t_{197.01}=-2.95$, $p = 0.040$, $\eta^2_p = 0.04$). Further, all except the $\beta_1$ band showed significance for the contrasts (1-3): $\beta_2$ ($t_{197.78} = -4.26$, $p = 0.001$, $\eta^2_p = 0.08$); $\gamma_1$ ($t_{197.80}=-4.00$, $p = 0.002$, $\eta^2_p = 0.07$); and $\gamma_2$ ($t_{197.82}=-3.71$, $p = 0.005$, $\eta^2_p = 0.07$). 

These findings suggest a strong distinction among the larger separation of workload levels for the Chess task in the high frequency bands. With increased power in the $\beta$ and $\gamma$ frequencies generally associated with active cognitive processing, attention, decision making, and mental effort \cite{jia2011gamma, KROPOTOV2016107}, the progressive increase in high-frequency band power from lower to higher workload conditions supports the interpretation that participants were engaging more cognitive resources as task complexity increased, particularly when comparing across 2 or 3 difficulty levels. 

\subsubsection{N-Back} 

In the N-Back task we observed significant workload effects in the lower frequency bands:  $\theta$ ($F_{3,141.15} = 4.90$, $p = 0.010$, $\eta^2_p = 0.09$), $\alpha_1$ ($F_{3,107.86} = 6.49$, $p = 0.003$, $\eta^2_p = 0.15$), and $\alpha_2$ ($F_{3,147.00} = 4.41$, $p = 0.012$, $\eta^2_p = 0.08$). Similar to the differentiations seen with the Chess task, post-hoc analyses (see \cref{supp-tab:WITHIN_EEG_PHOCS_NBACK} in the supplementary material) revealed the strongest effect between levels 0-3, represented in the $\theta$ ($t_{112.79} = -3.09$, $p < 0.048$, $\eta^2_p = 0.08$),  $\alpha_1$ ($t_{111.28} = -3.22$, $p < 0.048$, $\eta^2_p = 0.09$), and $\alpha_2$ ($t_{112.79} = -3.14$, $p = 0.048$, $\eta^2_p = 0.08$) frequencies. Effects between workload levels 0-2 were also represented in two out of three bands: $\theta$ ($t_{112.63}=-3.20$, $p = 0.048$, $\eta^2_p = 0.08$), and $\alpha_1$ ($t_{111.20}=-3.42$, $p = 0.048$, $\eta^2_p = 0.10$). 

These findings suggest a distinct pattern in the N-Back task with workload effects primarily manifesting in lower frequency bands. This directly confirms other EEG work done with the N-Back task, where an increase in the alpha and theta bands has been found between 0-Back and 2-Back conditions \cite{Brouwer2012}. The prominence of theta and alpha band activity further aligns with established research linking these frequencies to working memory operations, particularly the theta band's association with repetitive task load \cite{KROPOTOV200977}, and the alpha band's association selective attention, inhibition, and controlled access to stored information \cite{klimesch2012alpha}. Similar to the Chess task, the clearest differentiation between workload levels for the N-Back task showed similar effect sizes between the largest workload contrasts, suggesting that the subtlest distinctions in workload are not visible in the data. The overall pattern of increased lower frequency power with heightened working memory load reflects the task's core demand on memory maintenance and manipulation processes, distinguishing it from the more complex cognitive processing demands observed in the Chess task.

\subsubsection{Stroop}
Refer to \cref{supp-tab:WITHIN_EEG_ANOVAS} in the supplementary material. In the Stroop task we found significant distinction between congruent and incongruent stimuli (0-1) in the $\beta_2$ ($F_{1,44.52} = 9.32$, $p = 0.027$, $\eta^2_p = 0.17$) band only. However, $\alpha_2$ showed near significance with a high effect size ($F_{1,34.63} = 6.58$, $p = 0.052$, $\eta^2_p = 0.16$); we therefore ran post hoc tests on both. This revealed significance for both, showing $\alpha_2$ increasing during the incongruent task ($t_{40.00} = -2.57$, $p < 0.014$, $\eta^2_p = 0.14$), and $\beta_2$ decreasing during incongruent stimuli ($t_{40.00} = 2.57$, $p < 0.014$, $\eta^2_p = 0.14$).

Previous work has linked the Stroop task to a decline in beta band power, which may reflect cognitive control mechanisms involved in conflict resolution \cite{ZHAO2015130}. Supporting this finding, \cite{TAFURO2019107190} observed a similar decrease in $\beta$ band power during Stroop task performance, suggesting a consistent neural signature of conflict processing. The combination of this observation, along with an increase in $\alpha$ power consistent with the inhibitory mechanisms previously described, suggests a coordinated neural response during conflict resolution in the Stroop task.

\subsubsection{Rotation}
Notably, our analysis revealed no significant within-task workload differences across difficulty levels in the Rotation task. This consistent pattern of neural engagement suggests that cognitive demands as measured in the prefrontal area may not increase linearly with rotation angles, indicating that participants likely engaged similar neural mechanisms regardless of the rotation magnitude required. It is also possible, however, that the objectivity of the rating scale used may have confounded `true' difficulty. In \cite{so2017evaluation}'s similar work, difficulty levels were instead deduced from subjective ratings - it is possible that such difficulty scales might yield better results for this task. Further, analysis of neural signatures from the posterior regions may elicit more differentiable patterns. 

\subsection{Statistical Analysis of Cross-Task EEG Signal Differentiation}
ANOVA results (Table \cref{supp-tab:BETWEEN_EEG_ANOVAS}) revealed significant differences between tasks across all frequency bands ($F_{3,102.42-106.62} = 6.86-26.08$, all $p < 0.001$, $\eta^2_p=0.17-0.43$), with the strongest effects observed in lower frequency ranges: $\alpha_1$ ($\eta^2_p = 0.43$), $\alpha_2$ ($\eta^2_p = 0.37$), and $\theta$ ($\eta^2_p = 0.36$). Post-hoc analyses (\cref{supp-tab:BETWEEN_TASK_PHOCS} in the supplementary material and Figure \ref{fig:CROSS_TASK_RESULTS}) revealed distinct patterns across tasks which are discussed below.

\nopagebreak[1]
\begin{figure}[h!]
    \centering
    \includegraphics[width=\linewidth]{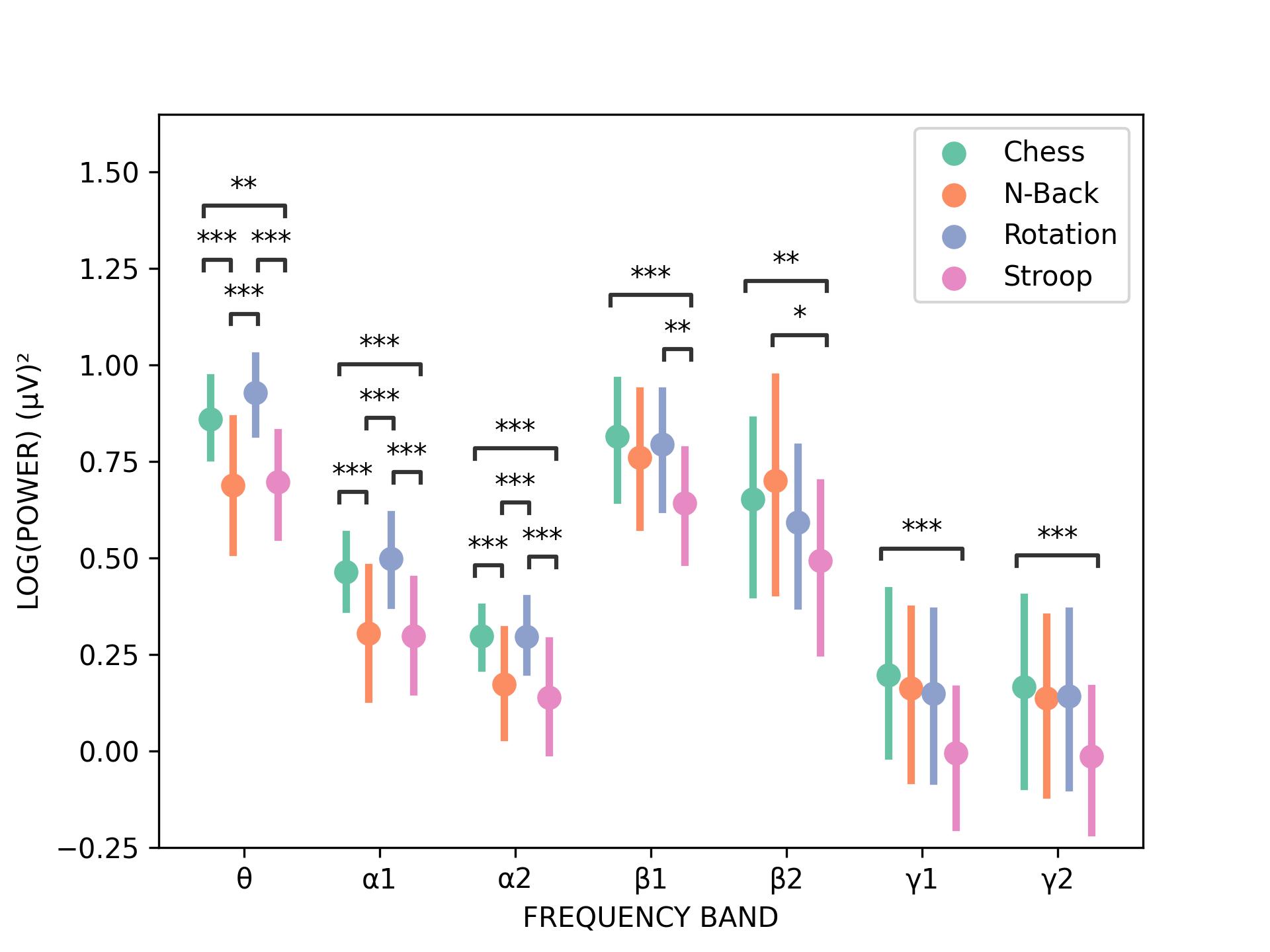}
    \caption{EEG spectral power compared across tasks, irrespective of workload level within each task. Differences in spectral power were observed across all frequency bands. Most notably, Chess shows significantly higher power than Stroop across all bands. Rotation likewise shows higher power than Stroop across multiple bands, but not in significantly in the high-frequency ranges. Both Chess and Rotation show higher power than N-Back in the lower frequency bands.}
    \label{fig:CROSS_TASK_RESULTS}
\end{figure}

From these data a few notable trends arise. Firstly are the similarities within the power disparities in the low-to-middle frequency bands ($\theta$, $\alpha_1$, and $\alpha_2$) between two groups of two tasks: Chess and Rotation, both significantly higher in all such bands (see \cref{supp-tab:BETWEEN_TASK_PHOCS} in the supplementary material) than the N-Back and Stroop tasks. This clustering suggests these tasks may share underlying cognitive mechanisms, particularly in how they engage spatial processing and mental manipulation of visual information. 

Considering this more deeply, as compared to Stroop and N-Back, both Chess and mental rotation heavily recruit visuospatial working memory systems. Although increases in $\alpha$ power have been found in relation to event-related desynchronization (ERS), other work has demonstrated that under certain conditions $\alpha$ power increases during task activation \cite{Klimesch2007}, specifically in a visuospatial manipulation task as compared to a memory task \cite{Sauseng2005}. Further, frontal $\alpha$ power has been linked to creative ideation reflecting internal processing \cite{Fink2014}, which largely would be exhibited during the Chess task. 

Elevated $\theta$ power likely reflects an increase of overall mental effort \cite{Smit2005} and on executive control processes \cite{Cavanagh2014, Sauseng2010} required by these tasks during the manipulation of complex spatial representations. Chess positions and mental rotation problems both require the generation and manipulation of spatial configurations within working memory, processes known to engage frontoparietal networks that communicate through synchronized $\theta$ oscillations \cite{Sauseng2010}. In the case of Chess puzzles, long-term memory structures will be engaged in the attempt to recall puzzle patterns previously encountered; in the case of mental rotation, as the visual attention is limited to a single block at a time, there is an implicit need to store and remember the block not currently attended to. Although the increase in $\theta$ power during increased levels of the N-Back task indicate increased storage and recall of information, the relative requirement of overall level-independent (i.e. all levels of N-Back) storage and recall is less than the additional neural processes required by Chess and Rotation, including manipulation of relatively larger quantities of visual information, and long-term informational interfacing in the case of Chess puzzles \cite{Sauseng2010}. Indeed, N-Back and Stroop tasks involve rapid (1s) sequential trials, which can provide targeted tasks to explore specific neurophysiological dynamics, but overall may involve less cognitive power than the activity of the Chess and Rotation tasks. 

This distinction also maps to Baddeley's multicomponent model of working memory \cite{Baddeley2000}: Chess and Rotation more heavily engage the visuospatial sketchpad component, whereas N-Back and Stroop by contrast more heavily engage the phonological loop.

The pattern between Chess, Rotation, and Stroop persists in the $\beta_1$ band, with a notable exception: the N-Back task is not significantly lower than either Chess or Rotation. This divergence likely stems from the N-Back task's intensive working memory demands, which have been shown to relate to bursts of activity in $\beta$ band power \cite{Lundqvist2018}. 

Moving into the higher frequency bands, N-Back and Chess show higher power than Stroop in the $\beta_2$ band. This result may potentially be a consequence of the decreased $\beta_2$ activation due to conflict inhibition discussed in the within-task results for Stroop, combined with increased mental workload exhibited by N-Back. Chess continues to have significantly higher power than Stroop in the highest bands ($\beta_2$, $\gamma_1$, and $\gamma_2$). This comparison is likely due to the engagement of integrated and complex cognitive processes in the Chess task \cite{Atherton2003}. Yet, although Chess represents the consistently highest total amount of $\beta$ and $\gamma$, no significant differences among Chess, N-Back, and Rotation were discovered in these bands, indicating that identifying the particular nuances of the differential nature of the complexity of the Chess task is more challenging. Inspection of distinctions among more areas of the brain may yield insights into the full complexities of Chess as compared to Rotation and N-Back.

Taken as a whole, these findings demonstrate the capability of low-cost EEG systems to differentiate between complex neural activation patterns across varied cognitive tasks, using only the prefrontal region. These results suggest promising applications for cognitive task classification and analysis.

\subsection{Within-Task Workload Machine Learning Results}
As shown in Table \ref{tab:ML_WITHIN_TASK}, the within-task workload classification presented significant challenges. The N-Back task achieved a 63\% F1-score, suggesting some promise, while other tasks performed near chance levels when using the MUSE 2 device for online classification. Several factors may explain these results. First, our limited dataset size may have constrained the models' learning capabilities. 

Secondly, the N-Back task inherently induces the most distinct cognitive load differences between difficulty levels \cite{Brouwer2012}. By contrast, although the Stroop task can reliably induce cognitive interference measurable at the macro level, the subtleties of the neural signatures elicited between the congruent and incongruent states may be  difficult to capture on a per-trial basis with the resolution of the MUSE 2 device \cite{krigolson2017choosing}. Likewise, although larger degrees of mental rotation take more time on average to complete, the neural signatures underlying the tasks may not be sufficiently distinct to measure with the MUSE 2. In the case of both Stroop and rotation, it is also possible that complex networks extending beyond the prefrontal region (in particular, parietal and visuospatial networks) may be required to interpret the data more clearly. Regarding Chess puzzles, it is possible that at the higher degrees of difficulty, time pressure may have introduced artifacts into the neural signals which might have affected classification performance.

Additionally, classification accuracy might improve through either the incorporation of additional features or the use of alternative models that process raw data directly rather than using frequency-domain transformations. These results demonstrate promising potential for applying the MUSE 2 device and similar consumer-grade EEG systems to distinguish between high and low workload levels as defined by the working-memory based cognitive load induced by the N-Back task. With additional data collection and refinement, this capability could potentially extend to Chess-based cognitive load classification. However, our findings suggest that the MUSE 2 may not be suitable for BCI applications that rely on cognitive inhibition or mental rotation tasks, and we recommend careful consideration before deploying it in such contexts.

\begin{table}[h!]
\centering
\caption{Task-specific classification of cognitive workload using machine learning. N-Back performs the best, with 63\% F1-score over 2-class classification. Chess performs just better than chance, with 54\%. The Rotation and Stroop task classifiers do not perform better than chance. Note that the results are averages over all Monte-Carlo iterations. }
\label{tab:ML_WITHIN_TASK}
\begin{tabular}{ccccccc}
\toprule
Task & Workload & Precision & Recall & F1 & Support \\
\midrule
\multirow{3}{*}{Chess} 
& 0 & 0.540 & 0.553 & 0.543 & 994\\
& 1 & 0.540 & 0.522 & 0.528 & 994\\
\cmidrule{2-6}
& Macro Average & 0.539 & 0.538 & 0.535 & 1987\\
\midrule
\multirow{3}{*}{N-Back} 
& 0 & 0.622 & 0.662 & 0.641 & 311 \\
& 1 & 0.638 & 0.596 & 0.616 & 311 \\
\cmidrule{2-6}
& Macro Average & 0.630 & 0.629 & 0.628 & 622 \\
\midrule
\multirow{3}{*}{Rotation} 
& 0 & 0.492 & 0.513 & 0.501 & 110 \\
& 1 & 0.492 & 0.471 & 0.481 & 110 \\
\cmidrule{2-6}
& Macro Average & 0.492 & 0.492 & 0.491 & 220\\
\midrule
\multirow{3}{*}{Stroop} 
& 0 & 0.507 & 0.572 & 0.538 & 202\\
& 1 & 0.509 & 0.443 & 0.474 & 202\\
\cmidrule{2-6}
& Macro Average & 0.508 & 0.508 & 0.506 & 404 \\
\bottomrule
\end{tabular}
\end{table}

\subsection{Cross-Task Machine Learning Results}
Full results are shown in Table \ref{tab:ML_CROSS_TASK}. The machine learning model for between-task classification demonstrated robust performance, achieving a macro-average F1-score of 49\% (95\% confidence interval of [0.493, 0.495]) compared to the expected random chance performance of 25\%. The Rotation task showed the strongest classification metric with an F1-score of 53\%, while Chess, Stroop, and N-Back tasks achieved F1-scores of 50\%, 49\%, and 45\%, respectively. This performance is particularly noteworthy given the inherent challenges of EEG classification. The balanced precision and recall scores across all tasks (ranging from approximately 46\% to 56\%) indicate consistent and reliable classification capabilities. The relatively similar performance levels of the tasks align with our earlier findings regarding unique combinations of neural activation patterns discussed earlier.
These classification results demonstrate that the MUSE 2 device can effectively differentiate between distinct cognitive tasks, with performance approximately twice that of random chance across all task types.

\begin{table}[h!]
\centering
\caption{4-way cross-task classification (random chance level is 25\%) of EEG signals using machine learning. These results are averages across all  Monte Carlo iterations.}
\label{tab:ML_CROSS_TASK}
\begin{tabular}{lcccc}
\toprule
Task & Precision & Recall & F1 & Support \\
\midrule
Chess & 0.552 & 0.460 & 0.499 & 195 \\
N-Back & 0.485 & 0.425 & 0.452 & 195 \\
Rotation & 0.493 & 0.580 & 0.531 & 195 \\
Stroop & 0.474 & 0.519 & 0.494 & 195 \\
\midrule
Macro Average & 0.501 & 0.496 & 0.494 & 780 \\
\bottomrule
\end{tabular}
\end{table}

\section{Future Work and Methodological Considerations}
\subsection{Signal Processing Enhancements}
Our current epoching and frequency-domain transformation approach, while effective for real-time applications, could be complemented by alternative analytical methods. Event-related potential (ERP) analysis would enable precise examination of temporal relationships between task events and neural responses, potentially revealing workload-sensitive components like the P300 or N400. This approach would be particularly valuable for the Stroop and Mental Rotation tasks, where specific cognitive processes occur in response to stimulus presentation. Advanced time-frequency methods, such as wavelet transforms or variable-window short-time Fourier transforms, could provide finer temporal resolution at higher frequencies while maintaining adequate frequency resolution for lower bands. These methods could capture subtle workload-related changes and transient spectral patterns that may be averaged out in our fixed-window approach. These might be combined with longer-epoch windows, enabling finer-grained insights into the temporal dynamics of cognitive workload across different task phases.

\subsection{Advanced Machine Learning Applications}
Recent developments in deep learning architectures suggest promising directions for improving workload classification. \cite{craik2019deep} demonstrates deep learning models' success in handling temporal dynamics inherent in EEG data, while \cite{hossain2023status} shows transformer-based architectures can effectively capture long-range dependencies in EEG signals. Building on methodological insights from \cite{murungi2023empowering}, future work could implement attention mechanisms and transfer learning to address our dataset limitations, particularly beneficial for complex tasks like Chess where traditional frequency-domain features may not fully capture neural dynamics.
Future investigations should balance architectural sophistication with practical implementation constraints for consumer-grade EEG devices. This aligns with current trends toward deployable BCI systems that maintain both accuracy and computational efficiency \cite{hossain2023status}. Additionally, incorporating self-supervised learning techniques could address data limitation challenges \cite{craik2019deep}, potentially improving classification robustness across cognitive tasks through more effective utilization of unlabeled EEG data.

\subsection{Sample Size and Generalizability}
The sample size of our dataset is small (22 participants total across Phase I and II), and the gender distribution is uneven (Phase II includes only male participants). These factors may affect the generalizability of the results. For future work, we recommend increasing the sample size, and/or balancing the gender ratio. Further, the participant pool is primarily students. Thus, further work should also be done with participants across different age groups. 

\subsection{Task Difficulty Calibration}
Notably, this work did not include subjective workload assessments (e.g. NASA-TLX), which may allow for more fine-grained calibration of tasks based on workload. Although in the current study paradigm it would be difficult to apply TLX to, for instance, levels of rotation or levels of difficulty during Chess puzzle playing, augmented designs with different levels of rotation as separate blocks may enable this kind of classification. 

\subsection{Probe Limitations}
Our study only focused on the frontal electrodes of the MUSE 2, at positions AF7 and AF8, which may miss critical signals from posterior brain regions (e.g. parietal areas). Further work focusing on introducing low-cost sensors with better spatial resolution would likely improve classification accuracy and comprehensiveness of the results.  

\subsection{Signal Connectivity Issues}
Data loss due to wireless connectivity issues posed a significant limitation on our ability to collect data. Wired connections for the future would be beneficial, or improved algorithms to monitor real-time quality of the signal. 


\section{Conclusion}
Our study presents a novel chess puzzles paradigm for cognitive neuroscience, provides an open-access framework to duplicate and modify the task paradigm including this task and 3 common cognitive psychology tasks within the context of the MUSE 2 device, provides an open-access dataset of neural data for researchers, and yielded several significant findings regarding the capabilities of the MUSE 2 device for cognitive analysis and brain-computer interface applications.

We demonstrated that the MUSE 2 can detect subtle variations in differing forms cognitive load across both traditional experimental paradigms (N-Back and Stroop tasks) and more ecologically valid scenarios (Chess puzzles task). The cross-task analysis revealed distinct neural signatures that show promise for BCI applications. Our Monte Carlo cross-validation procedure provided robust testing of the system's predictive capabilities for real-time BCI implementations. 

The within-task workload classification results indicate that while the MUSE 2 can effectively track cognitive load gradients in the N-Back task, it may have limitations for fine-grained distinctions within Chess, Rotation, or Stroop tasks for online classification. This suggests opportunities for future research exploring advanced preprocessing techniques and alternative machine learning approaches. 

Notably, our cross-task classification results demonstrate particularly promising applications for the MUSE 2 in BCI research. Rather than focusing on within-task workload assessment, the device shows significant potential for broader task differentiation, indicating that real-time BCIs can be developed for meaningful human state classification. 

In conclusion, this research presents a novel Chess puzzle task, and establishes that consumer-grade EEG devices like the MUSE 2 can offer viable solutions for detecting distinctions in neurological states during various cognitive tasks, can be used to classify mental workload in the N-Back task, and that potential exists to differentiate among neurological patterns across diverse cognitive tasks. While challenges remain, these findings open up exciting possibilities for future development of adaptive brain-computer interfaces. Openly available experimental code and data can further empower future researchers to explore similar task paradigms. 

\section*{Data Availability}
The research code used to run the experimental tasks can be accessed at \url{REDACTED}. 


The preprocessed EEG datasets can be found at 
\url{REDACTED}. 
The chess puzzles database is not maintained by the study team, and can instead be downloaded from \url{https://database.lichess.org/} \cite{lichess2024}.

\bibliographystyle{IEEEtran}
\bibliography{references}




\end{document}


\date{}
\title{Supplementary Material}
\maketitle
\newpage

\begin{figure}[h!]
\centering
\includegraphics[width=.8\linewidth]{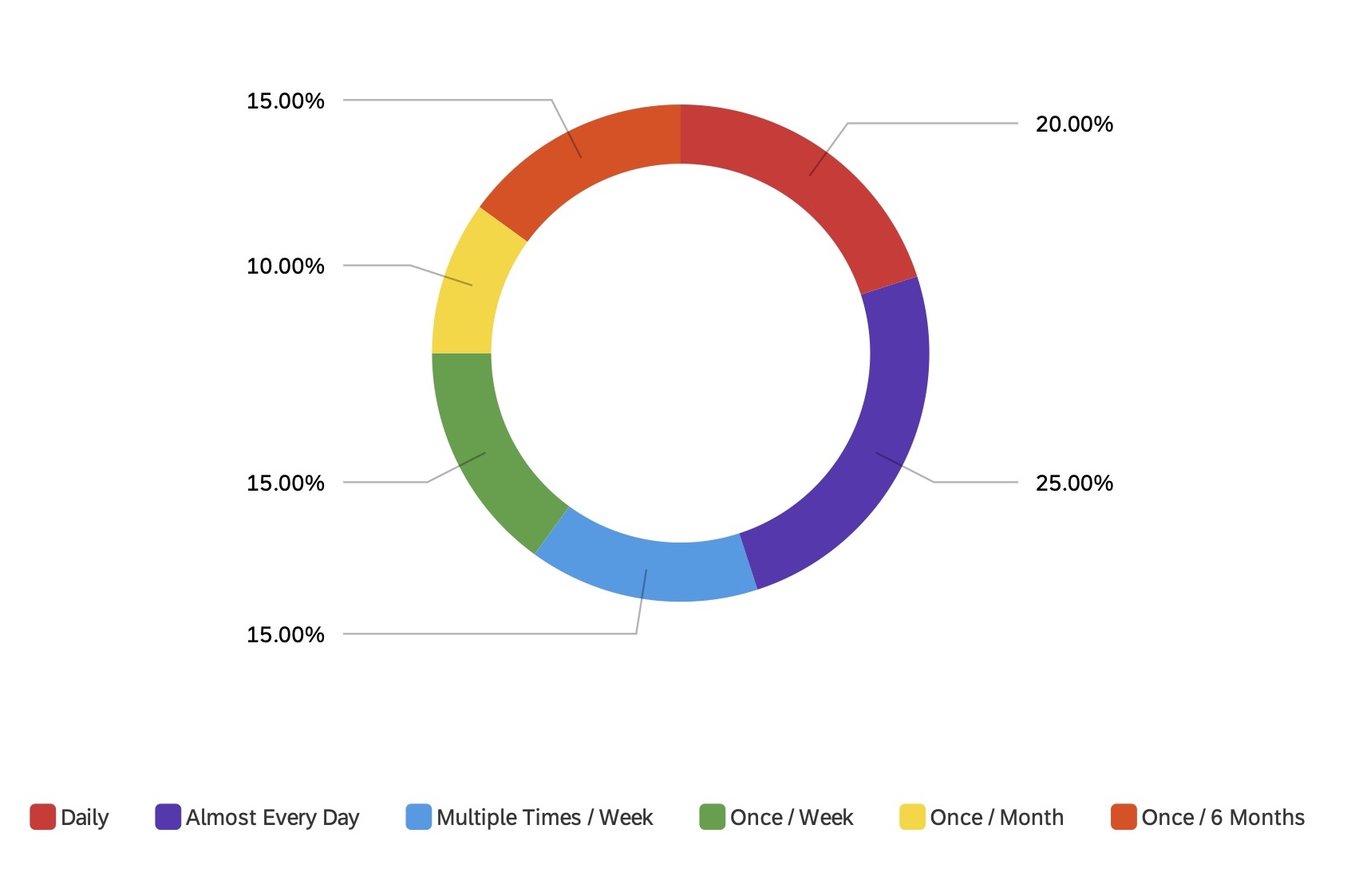}
\caption{Frequency of Chess play across all Chess participants (N=17). Most players (75\%) played Chess at least once per week. }
\label{fig:PLAY_FREQUENCY}
\end{figure}

\begin{figure}[h!]
\centering
\includegraphics[width=.8\linewidth]{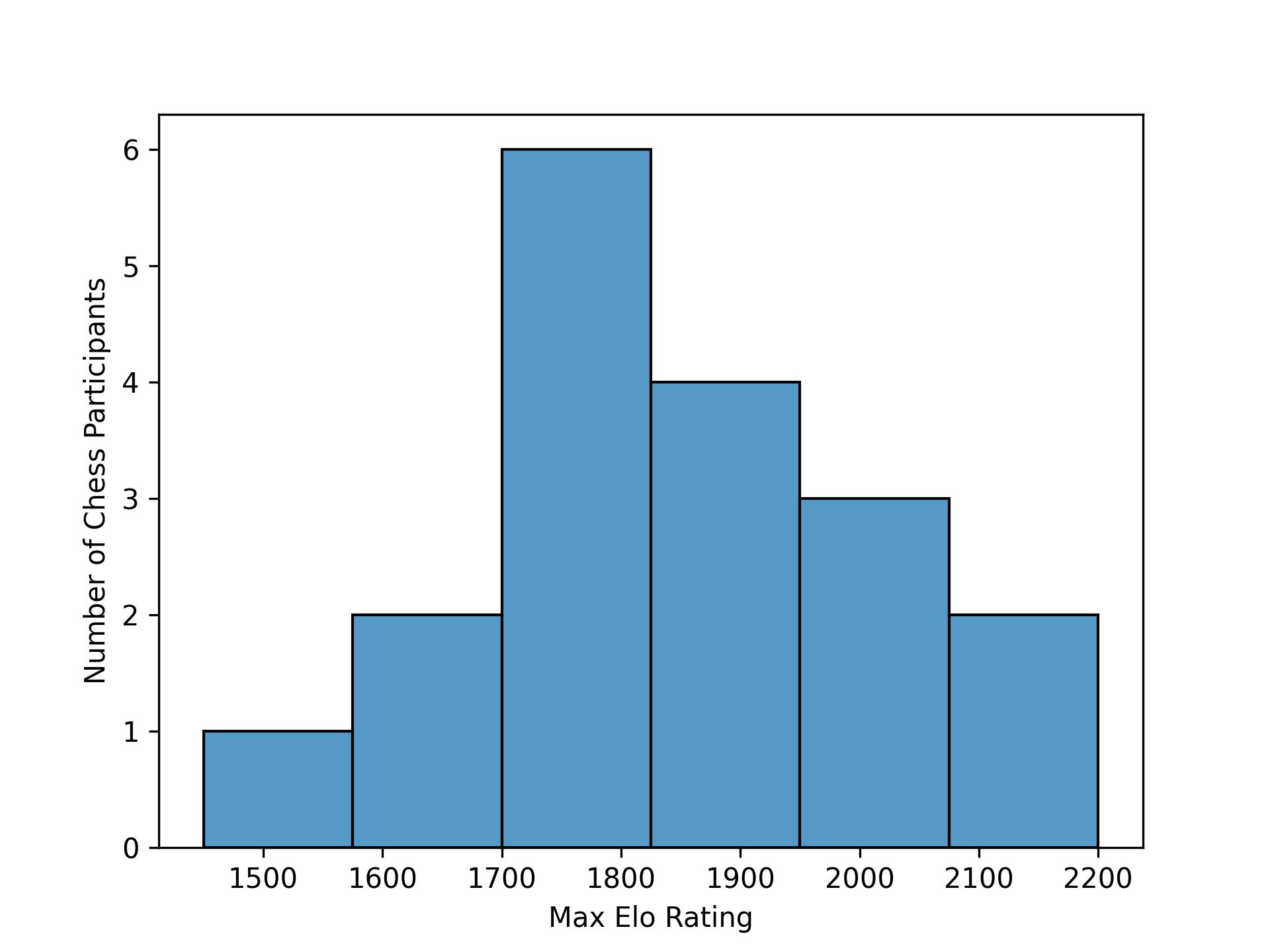}
\caption{Distribution of maximum Chess puzzle difficulty level achieved across participants, measured in Elo Glicko2 rating. Preliminary quantile-based analyses revealed no significant associations between maximum Elo rating and EEG response patterns (analyses not reported in main results).}
\label{fig:MAX_ELO}
\end{figure}

\begin{table}[h!]
\centering
\caption{Detailed dataset sizes after exclusions and filtering. Data here are reported from both phases I and phase II. Features per Task, are the number of participants and (Num. P) mean Blocks per Participant (B/P). For each level of Workload (W-Load) within Task, the features are: mean Trials per Block per Participant (T/B/P), total number of Trials (T), mean Epochs per Block per Participant (E/B/P), and total Epochs (E). Note that, although the number of trials is similar across workload levels, the total number of epochs increases as workload increases, because there will be more 1 second samples in higher workload conditions. For statistical modeling we average all epochs for each participant at the block level (and within levels of workload for within-task workload classification); for machine learning, epochs within each block, each of which contain all workload levels, are grouped together and used exclusively in the training or testing set. }
\label{tab:DATA_DESCRIPTION}
\begin{adjustbox}{width=\linewidth}
\begin{tabular}{ccc|ccccc}
\toprule
Task & Num. P & Mean B/P & W-Load & Mean T/B/P & Total T & Mean E/B/P & Total E \\
\midrule
\multirow{5}{*}{Chess} 
& && 0 & 5.7 & 556 & 45.3 &3109\\
& && 1  & 6.4 & 631 &68.2 &4690 \\
& 16& 5.4 & 2 & 10.2 & 989& 129.6 &8981\\
& && 3 & 8.3 & 758 &146.9& 9582\\
\cmidrule{4-8}
& && Total & 30.65 & 2934& 389.86 &26362\\
\midrule
\multirow{5}{*}{N-Back} 
& && 0 & 22.3 &1397& 27.0& 1058\\
& && 1  & 22.4 &1345& 29.0& 1050 \\
&11&4 & 2 & 23.1  &1480 &33.7 &1328\\
& && 3 & 90.9 & 1438 &36.4 &1332\\
\cmidrule{4-8}
& && Total & 22.7 & 5660 &126.1&4768 \\
\midrule
\multirow{5}{*}{Rotation} 
& && 0 & 5.8 & 388&11.1& 433\\
& && 1  & 6.1 &411 &14.2&  579\\
&11&3.9 & 2 & 6.2  & 409 &17.2& 717\\
& && 3 & 5.8 & 387 &16.7& 715\\
\cmidrule{4-8}
& && Total & 23.9 & 1595 &59.2&2444 \\
\midrule
\multirow{3}{*}{Stroop} 
&& & 0 & 33.9 & 2260 &31.9& 1302\\
&10&3.9 & 1 & 35.2 & 2317 &36.6&  1391\\
\cmidrule{4-8}
& && Total & 69.1 &4577 &68.6& 2793\\
\bottomrule
\end{tabular}
\end{adjustbox}
\end{table}

\begin{figure}[h!]
\centering
\includegraphics[width=\linewidth]{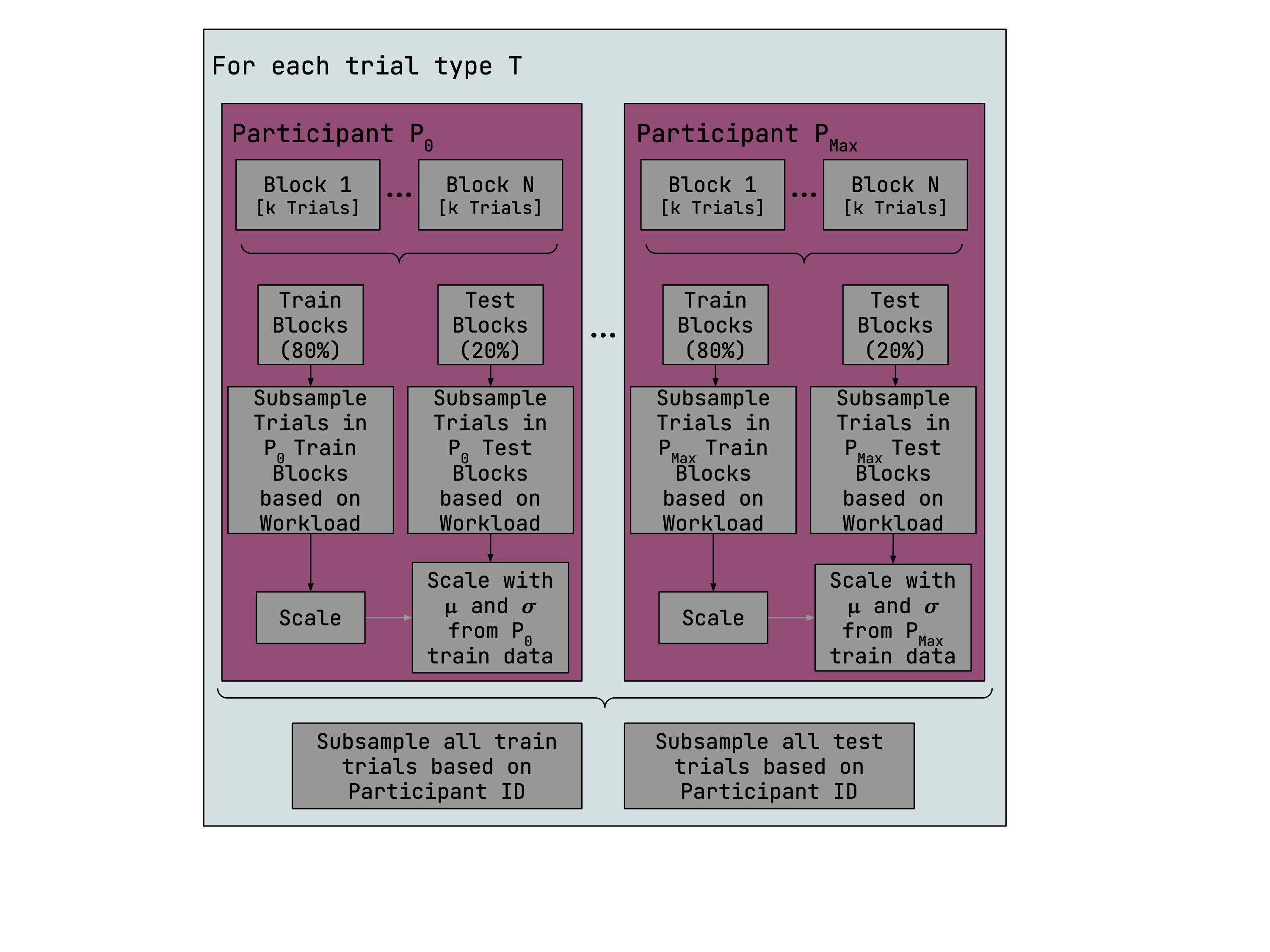}
    \caption{Visualization of one iteration of the data splitting steps for the Monte Carlo pipeline for within-task workload classification. Blocks of grouped trials for each participant are split into training and testing sets randomly, with 80\% of blocks in training and the remaining 20\% of blocks in testing. Trials within training and testing blocks are subsampled randomly to the lowest frequency of workload label. Training data are scaled, and $\mu$ and $\sigma$ from the training data are used to scale the test data for that participant. All participants' training sets are then combined and subsampled to the lowest frequency participant identifier; likewise happens for the test sets. This pipeline is done separately for each Monte Carlo simulation for each trial type. 
    }
    \label{fig:ML_PIPELINE_WITHIN_TASK_WORKLOAD}
\end{figure}

\begin{figure}[h!]
\centering
\includegraphics[width=.9\linewidth]{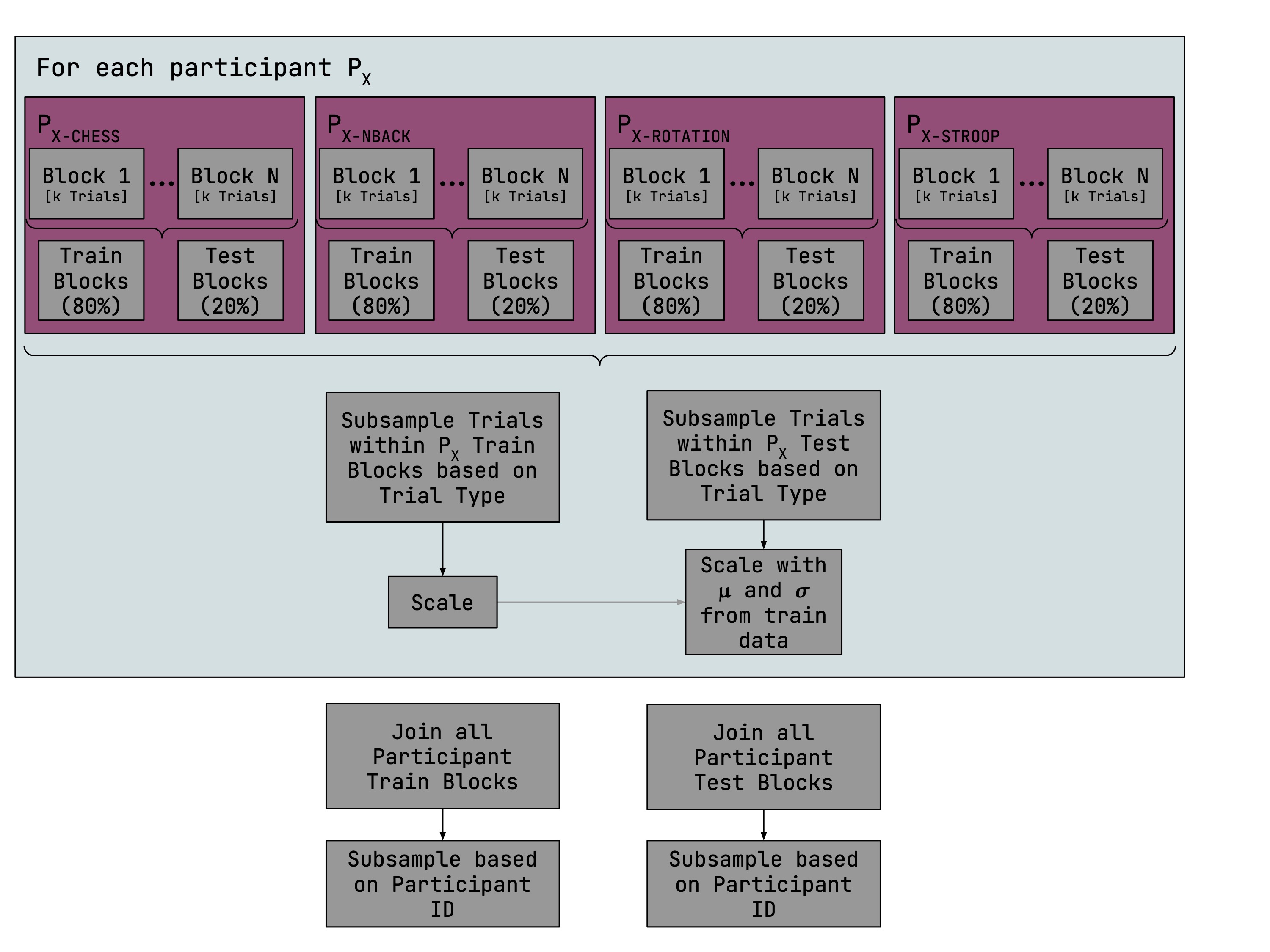}
    \caption{Visualization of one iteration of the data splitting steps for the Monte Carlo pipeline for cross-task classification. Blocks of grouped trials within each trial type for each participant are split into training and testing blocks randomly, with 80\% of blocks in training and the remaining 20\% of blocks in testing. All such training and testing blocks are combined for a given participant, and trials within the training and testing blocks are then subsampled (separately) to the lowest frequency label, and scaled. All such training and testing blocks for all participants and trial types are combined into single train and test sets, which are separately subsampled to the lowest frequency participant ID. F1 results are reported from both the precision and recall scores on the test set overall, and per-participant. }
    \label{fig:ML_PIPELINE_CROSS_TASK}
\end{figure}

\begin{table}[h!]
\caption{Number of participants per-model, and average train and test set sizes (rounded down) over all Monte Carlo iterations of within-task workload machine learning analysis. The data splitting procedure used ensures that for both test and train sets the labels in the model are equally balanced for each participant. Participants are likewise equally represented in both train and test sets.}
\centering
\label{tab:PPM_WORKLOAD}
\begin{adjustbox}{width=\linewidth}
\begin{tabular}{lccc}
\toprule
 Trial type & Number of participants & Average train set sample size & Average test set sample size \\
\midrule
Chess & 17 & 3,983 & 1987 \\
N-Back & 11 & 1,964 & 622 \\
Rotation & 10 & 842 & 220 \\
Stroop & 10 & 1,476 & 404 \\
\bottomrule
\end{tabular}
\end{adjustbox}
\end{table}

\begin{figure}[h!]
    \centering
    \includegraphics[width=\linewidth]{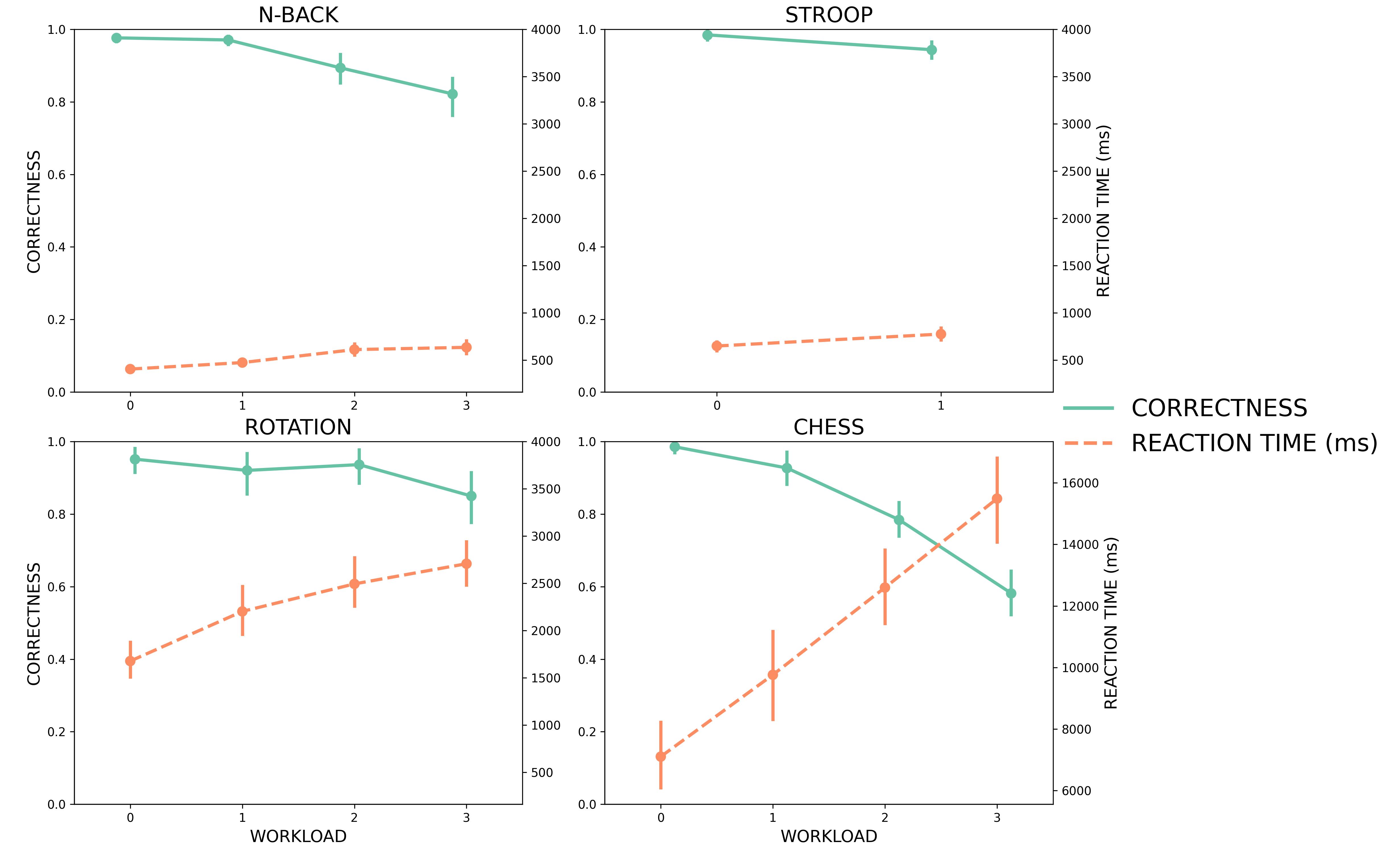}
    \caption{Performance and reaction time as a function of workload level for each of the four tasks. In general, as workload increases, correctness decreases and reaction time increases. Note that the temporal (right-side) scale y-scales for the three cognitive neuroscience tasks are the same (0-4000ms), however the same axis for the Chess task is from (6000-17000ms).  }
    \label{fig:BEHAVIORAL_STATS}
\end{figure}

\begin{table}[h!]
\centering
\caption{Post-hoc contrast results for modeling reaction time as a function of workload level for each task. Separate models were created for each task type; p-values are corrected using the Benjamini-Hochberg procedure. Outside of levels 2-3 in the N-Back task, reaction time increased significantly across difficulty levels of all tasks.} 
\label{tab:BEHAVIORAL_TIME_POST_HOC}
\begin{adjustbox}{width=\linewidth}
\begin{tabular}{lcccccccccr}
\toprule
Task & Contrast & Est. & SE & df & t & p & p.adj & p.adj.sig & $\eta_{p}^{2}$ & 95\% CI \\
\midrule

Chess & 0 - 3 & -0.34 & 0.01 & 2913.17 & -27.70 & $<$0.001 & $<$0.001 & *** & 0.21 & [0.19,1] \\
Chess & 0 - 2 & -0.23 & 0.01 & 2886.14 & -20.14 & $<$0.001 & $<$0.001 & *** & 0.12 & [0.11,1] \\
Chess & 1 - 3 & -0.23 & 0.01 & 2911.83 & -19.71 & $<$0.001 & $<$0.001 & *** & 0.12 & [0.10,1] \\
Chess & 0 - 1 & -0.10 & 0.01 & 2868.08 & -8.26 & $<$0.001 & $<$0.001 & *** & 0.02 & [0.02,1] \\
Chess & 1 - 2 & -0.13 & 0.01 & 2909.87 & -11.55 & $<$0.001 & $<$0.001 & *** & 0.04 & [0.03,1] \\
Chess & 2 - 3 & -0.10 & 0.01 & 2821.98 & -9.82 & $<$0.001 & $<$0.001 & *** & 0.03 & [0.02,1] \\
\midrule
N-Back & 0 - 3 & -0.17 & 0.00 & 6118.84 & -41.79 & $<$0.001 & $<$0.001 & *** & 0.22 & [0.21,1] \\
N-Back & 0 - 2 & -0.16 & 0.00 & 6124.60 & -40.03 & $<$0.001 & $<$0.001 & *** & 0.21 & [0.19,1] \\
N-Back & 1 - 3 & -0.10 & 0.00 & 6111.06 & -24.76 & $<$0.001 & $<$0.001 & *** & 0.09 & [0.08,1] \\
N-Back & 0 - 1 & -0.07 & 0.00 & 6115.00 & -16.82 & $<$0.001 & $<$0.001 & *** & 0.04 & [0.04,1] \\
N-Back & 1 - 2 & -0.09 & 0.00 & 6117.60 & -22.74 & $<$0.001 & $<$0.001 & *** & 0.08 & [0.07,1] \\
N-Back & 2 - 3 & -0.01 & 0.00 & 6114.86 & -2.18 & 0.128 & 0.128 & ns & 0.00 & [0.00,1] \\
\midrule
Stroop & 0 - 1 & -0.07 & 0.00 & 4894.41 & -22.83 & $<$0.001 & $<$0.001 & *** & 0.10 & [0.08,1] \\
\midrule
Rotation & 0 - 3 & -0.23 & 0.01 & 1552.22 & -18.09 & $<$0.001 & $<$0.001 & *** & 0.17 & [0.15,1] \\
Rotation & 0 - 2 & -0.19 & 0.01 & 1557.22 & -15.37 & $<$0.001 & $<$0.001 & *** & 0.13 & [0.11,1] \\
Rotation & 1 - 3 & -0.11 & 0.01 & 1558.35 & -8.51 & $<$0.001 & $<$0.001 & *** & 0.04 & [0.03,1] \\
Rotation & 0 - 1 & -0.12 & 0.01 & 1548.43 & -9.85 & $<$0.001 & $<$0.001 & *** & 0.06 & [0.04,1] \\
Rotation & 1 - 2 & -0.07 & 0.01 & 1548.64 & -5.70 & $<$0.001 & $<$0.001 & *** & 0.02 & [0.01,1] \\
Rotation & 2 - 3 & -0.04 & 0.01 & 1562.05 & -2.90 & 0.020 & 0.021 & * & 0.01 & [0.00,1] \\

\bottomrule
\end{tabular}
\end{adjustbox}
\end{table}

\begin{table}[h!]
\centering
\caption{Post-hoc contrast results for modeling correctness as a function of workload level for each task. Separate models were created for each task type; p-values are corrected using the Benjamini-Hochberg procedure. Outside of N-Back levels 0-1 and differences in Rotation difficulty among 0, 1, and 2, workload level significantly predicts correctness in differences for all tasks.}
\label{tab:BEHAVIORAL_CORRECTNESS_POST_HOC}
\begin{adjustbox}{width=\linewidth}
\begin{tabular}{lcccccccc}
\toprule
Task & Contrast & Est. & SE & df & z & p & p.adj & p.adj.sig \\
\midrule
Chess & 0 - 3 & 3.96 & 0.37 & inf & 10.76 & $<$0.001 & $<$0.001 & *** \\
Chess & 0 - 2 & 2.86 & 0.36 & inf & 7.87 & $<$0.001 & $<$0.001 & *** \\
Chess & 1 - 3 & 2.33 & 0.19 & inf & 12.55 & $<$0.001 & $<$0.001 & *** \\
Chess & 0 - 1 & 1.62 & 0.39 & inf & 4.19 & $<$0.001 & $<$0.001 & *** \\
Chess & 1 - 2 & 1.24 & 0.17 & inf & 7.15 & $<$0.001 & $<$0.001 & *** \\
Chess & 2 - 3 & 1.09 & 0.13 & inf & 8.69 & $<$0.001 & $<$0.001 & *** \\
\midrule
N-Back & 0 - 3 & 2.28 & 0.18 & inf & 12.61 & $<$0.001 & $<$0.001 & *** \\
N-Back & 0 - 2 & 1.63 & 0.19 & inf & 8.74 & $<$0.001 & $<$0.001 & *** \\
N-Back & 1 - 3 & 2.06 & 0.17 & inf & 12.19 & $<$0.001 & $<$0.001 & *** \\
N-Back & 0 - 1 & 0.22 & 0.23 & inf & 0.97 & 0.765 & 0.765 & ns \\
N-Back & 1 - 2 & 1.41 & 0.18 & inf & 8.05 & $<$0.001 & $<$0.001 & *** \\
N-Back & 2 - 3 & 0.66 & 0.11 & inf & 6.02 & $<$0.001 & $<$0.001 & *** \\
\midrule
Stroop & 0 - 1 & 1.38 & 0.18 & inf & 7.50 & $<$0.001 & $<$0.001 & *** \\
\midrule
Rotation & 0 - 3 & 1.38 & 0.28 & inf & 4.88 & $<$0.001 & $<$0.001 & *** \\
Rotation & 0 - 2 & 0.36 & 0.32 & inf & 1.12 & 0.675 & 0.754 & ns \\
Rotation & 1 - 3 & 0.74 & 0.24 & inf & 3.09 & 0.011 & 0.014 & * \\
Rotation & 0 - 1 & 0.65 & 0.30 & inf & 2.12 & 0.145 & 0.172 & ns \\
Rotation & 1 - 2 & -0.29 & 0.28 & inf & -1.04 & 0.726 & 0.765 & ns \\
Rotation & 2 - 3 & 1.03 & 0.26 & inf & 4.03 & $<$0.001 & $<$0.001 & *** \\
\bottomrule
\end{tabular}
\end{adjustbox}
\end{table}

\begin{figure}
    \centering
    \includegraphics[width=\linewidth]{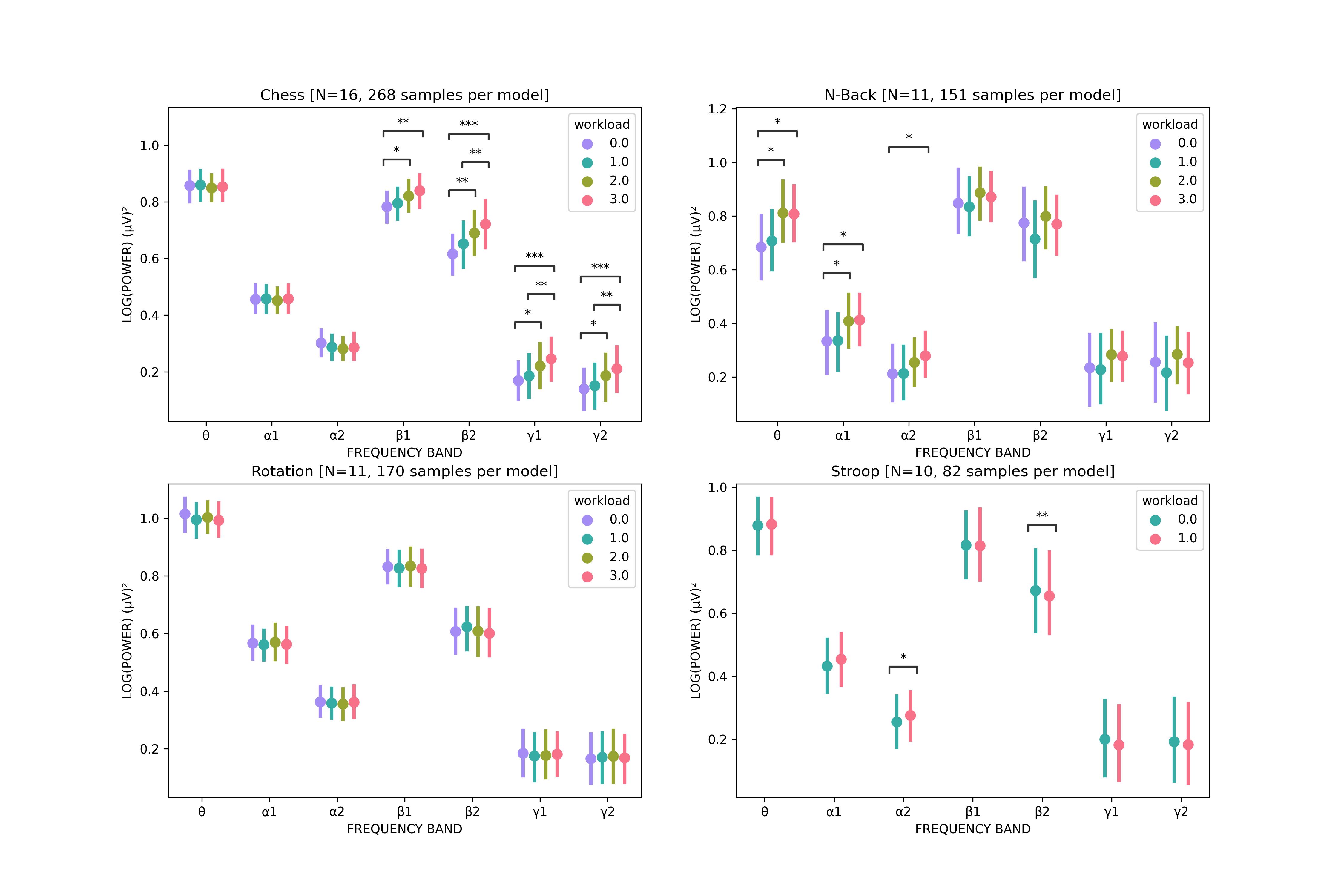}
    \caption{EEG power band data as compared across levels of workload within each task. Across workload levels Chess showed significant differences in high-frequency bands ($\beta_1$, $\beta_2$, $\gamma_1$, $\gamma_2$). Conversely, N-Back showed significant differences in the lower-frequency bands ($\theta$, $\alpha_1$, $\alpha_2$). Stroop showed significance in $\alpha_2$ and $\beta_2$, and Rotation did not show any significant differences across power bands. }
    \label{fig:WITHIN_TASK_WORKLOAD}
\end{figure}

\begin{table}[h!]
\centering
\caption{ANOVA results modeling EEG band data as a function of within-task workload. Separate LMER models were created for each unique combination of wavelength level and task. P-values are corrected with Benjamini-Hochberg procedure across all models. With increased workload levels, Chess showed changes in the upper range of frequency bands, and N-Back showed increased power in the lower range of frequency bands. Only $\beta_2$ was the only significant wavelength in the Stroop task, although there is a notably high effect size for $\alpha_2$. No changes were observed in the Rotation task. }
\label{tab:WITHIN_EEG_ANOVAS}
\begin{adjustbox}{width=\linewidth}
\begin{tabular}{lccccccccc}
\toprule
Task & Band & NumDF & DenDF & F & p & p.adj & p.adj.sig & $\eta_{p}^{2}$ & 95\% CI \\
\midrule
Chess & $\theta$ & 3.00 & 206.78 & 0.18 & 0.911 & 0.911 & ns & 0.00 & [0.00,1] \\
Chess & $\alpha_1$ & 3.00 & 207.36 & 0.35 & 0.791 & 0.911 & ns & 0.01 & [0.00,1] \\
Chess & $\alpha_2$ & 3.00 & 207.33 & 2.58 & 0.055 & 0.077 & ns & 0.04 & [0.00,1] \\
Chess & $\beta_1$ & 3.00 & 204.91 & 6.27 & $<$0.001 & 0.001 & *** & 0.08 & [0.03,1] \\
Chess & $\beta_2$ & 3.00 & 204.27 & 13.40 & $<$0.001 & $<$0.001 & *** & 0.16 & [0.09,1] \\
Chess & $\gamma_1$ & 3.00 & 205.27 & 10.29 & $<$0.001 & $<$0.001 & *** & 0.13 & [0.06,1] \\
Chess & $\gamma_2$ & 3.00 & 205.98 & 8.35 & $<$0.001 & $<$0.001 & *** & 0.11 & [0.04,1] \\
\midrule
N-Back & $\theta$ & 3.00 & 141.15 & 4.90 & 0.003 & 0.010 & * & 0.09 & [0.02,1] \\
N-Back & $\alpha_1$ & 3.00 & 107.86 & 6.49 & $<$0.001 & 0.003 & ** & 0.15 & [0.05,1] \\
N-Back & $\alpha_2$ & 3.00 & 147.00 & 4.41 & 0.005 & 0.012 & * & 0.08 & [0.02,1] \\
N-Back & $\beta_1$ & 3.00 & 112.76 & 2.13 & 0.100 & 0.154 & ns & 0.05 & [0.00,1] \\
N-Back & $\beta_2$ & 3.00 & 112.57 & 1.23 & 0.303 & 0.303 & ns & 0.03 & [0.00,1] \\
N-Back & $\gamma_1$ & 3.00 & 112.54 & 1.76 & 0.159 & 0.186 & ns & 0.04 & [0.00,1] \\
N-Back & $\gamma_2$ & 3.00 & 112.81 & 2.06 & 0.110 & 0.154 & ns & 0.05 & [0.00,1] \\
\midrule
Stroop & $\theta$ & 1.00 & 39.92 & 0.00 & 0.980 & 0.980 & ns & 0.00 & [0.00,1] \\
Stroop & $\alpha_1$ & 1.00 & 34.99 & 3.15 & 0.085 & 0.149 & ns & 0.08 & [0.00,1] \\
Stroop & $\alpha_2$ & 1.00 & 34.63 & 6.58 & 0.015 & 0.052 & ns & 0.16 & [0.02,1] \\
Stroop & $\beta_1$ & 1.00 & 44.05 & 0.08 & 0.772 & 0.901 & ns & 0.00 & [0.00,1] \\
Stroop & $\beta_2$ & 1.00 & 44.52 & 9.32 & 0.004 & 0.027 & * & 0.17 & [0.04,1] \\
Stroop & $\gamma_1$ & 1.00 & 40.78 & 4.43 & 0.042 & 0.097 & ns & 0.10 & [0.00,1] \\
Stroop & $\gamma_2$ & 1.00 & 44.28 & 0.17 & 0.681 & 0.901 & ns & 0.00 & [0.00,1] \\
\midrule
Rotation & $\theta$ & 3.00 & 129.00 & 0.86 & 0.465 & 0.926 & ns & 0.02 & [0.00,1] \\
Rotation & $\alpha_1$ & 3.00 & 126.15 & 0.17 & 0.917 & 0.926 & ns & 0.00 & [0.00,1] \\
Rotation & $\alpha_2$ & 3.00 & 113.15 & 0.19 & 0.901 & 0.926 & ns & 0.01 & [0.00,1] \\
Rotation & $\beta_1$ & 3.00 & 130.55 & 0.26 & 0.852 & 0.926 & ns & 0.01 & [0.00,1] \\
Rotation & $\beta_2$ & 3.00 & 130.25 & 1.31 & 0.272 & 0.926 & ns & 0.03 & [0.00,1] \\
Rotation & $\gamma_1$ & 3.00 & 130.29 & 0.34 & 0.796 & 0.926 & ns & 0.01 & [0.00,1] \\
Rotation & $\gamma_2$ & 3.00 & 130.49 & 0.15 & 0.926 & 0.926 & ns & 0.00 & [0.00,1] \\
\bottomrule
\end{tabular}
\end{adjustbox}
\end{table}

\begin{table}[h!]
\centering
\caption{ANOVA results from modeling EEG waveband data as a function of task type. To avoid multicollinearity between frequency bands, separate models were created for each waveband. P-values were adjusted using the Benjamini-Hochberg procedure to control for multiple comparisons. Significant differences in neural activity among tasks were observed across all frequency bands analyzed.}
\label{tab:BETWEEN_EEG_ANOVAS}
\begin{adjustbox}{width=\linewidth}
\begin{tabular}{lcccccccc}
\toprule
Band & NumDF & DenDF & F value & Pr($>$F) & p.adj & p.adj.sig & $\eta_{p}^{2}$ & 95\% CI \\
\midrule
$\theta$ & 3.00 & 101.52 & 18.35 & $<$0.001 & $<$0.001 & *** & 0.35 & [0.22,1] \\
$\alpha_1$ & 3.00 & 100.10 & 24.48 & $<$0.001 & $<$0.001 & *** & 0.42 & [0.30,1] \\
$\alpha_2$ & 3.00 & 98.74 & 18.82 & $<$0.001 & $<$0.001 & *** & 0.36 & [0.23,1] \\
$\beta_1$ & 3.00 & 97.21 & 9.25 & $<$0.001 & $<$0.001 & *** & 0.22 & [0.10,1] \\
$\beta_2$ & 3.00 & 97.09 & 6.88 & $<$0.001 & $<$0.001 & *** & 0.18 & [0.06,1] \\
$\gamma_1$ & 3.00 & 96.95 & 8.56 & $<$0.001 & $<$0.001 & *** & 0.21 & [0.09,1] \\
$\gamma_2$ & 3.00 & 97.20 & 6.34 & 0.001 & 0.001 & *** & 0.16 & [0.05,1] \\
\bottomrule
\end{tabular}
\end{adjustbox}
\end{table}

\begin{table}[h!]
\centering
\caption{Post-hoc contrast results from modeling EEG waveband data as a function of task. Significant differences were discovered across a variety of frequency bands for varying tasks, indicating that differing neural signatures associated with the tasks are measurable with the MUSE 2 device. Among wavebands, effect sizes in the relatively low bands ($\theta$, $\alpha_1$, $\alpha_2$) are the largest, with a notable exceptions of increased power of Chess as compared to Stroop across all bands.} 
\label{tab:BETWEEN_TASK_PHOCS}
\begin{adjustbox}{width=\linewidth}
\begin{tabular}{cllccccccccc}
\toprule
Band & Task1 & Task2 & Est. & SE & df & t & p & p.adj & p.adj.sig & $\eta_{p}^{2}$ & 95\% CI \\
\midrule
$\theta$ & Chess & N-Back & 0.18 & 0.04 & 100.35 & 4.91 & $<$0.001 & $<$0.001 & *** & 0.19 & [0.09,1] \\
$\theta$ & Chess & Rotation & -0.05 & 0.04 & 98.11 & -1.37 & 0.524 & 0.667 & ns & 0.02 & [0.00,1] \\
$\theta$ & Chess & Stroop & 0.18 & 0.04 & 112.30 & 4.40 & $<$0.001 & $<$0.001 & *** & 0.15 & [0.06,1] \\
$\theta$ & N-Back & Rotation & -0.23 & 0.04 & 98.07 & -6.19 & $<$0.001 & $<$0.001 & *** & 0.28 & [0.16,1] \\
$\theta$ & N-Back & Stroop & -0.00 & 0.04 & 112.11 & -0.12 & 0.999 & 0.999 & ns & 0.00 & [0.00,1] \\
$\theta$ & Rotation & Stroop & 0.23 & 0.04 & 109.75 & 5.72 & $<$0.001 & $<$0.001 & *** & 0.23 & [0.12,1] \\
\midrule
$\alpha_1$ & Chess & N-Back & 0.16 & 0.03 & 99.32 & 5.76 & $<$0.001 & $<$0.001 & *** & 0.25 & [0.14,1] \\
$\alpha_1$ & Chess & Rotation & -0.03 & 0.03 & 97.18 & -1.11 & 0.682 & 0.842 & ns & 0.01 & [0.00,1] \\
$\alpha_1$ & Chess & Stroop & 0.16 & 0.03 & 110.31 & 5.51 & $<$0.001 & $<$0.001 & *** & 0.22 & [0.11,1] \\
$\alpha_1$ & N-Back & Rotation & -0.19 & 0.03 & 98.16 & -6.78 & $<$0.001 & $<$0.001 & *** & 0.32 & [0.20,1] \\
$\alpha_1$ & N-Back & Stroop & 0.01 & 0.03 & 111.33 & 0.20 & 0.997 & 0.999 & ns & 0.00 & [0.00,1] \\
$\alpha_1$ & Rotation & Stroop & 0.19 & 0.03 & 109.25 & 6.61 & $<$0.001 & $<$0.001 & *** & 0.29 & [0.17,1] \\
\midrule
$\alpha_2$ & Chess & Stroop & 0.16 & 0.03 & 110.38 & 5.95 & $<$0.001 & $<$0.001 & *** & 0.24 & [0.14,1] \\
$\alpha_2$ & Chess & N-Back & 0.13 & 0.02 & 99.30 & 5.37 & $<$0.001 & $<$0.001 & *** & 0.23 & [0.12,1] \\
$\alpha_2$ & Chess & Rotation & 0.01 & 0.03 & 97.14 & 0.39 & 0.979 & 0.999 & ns & 0.00 & [0.00,1] \\
$\alpha_2$ & N-Back & Rotation & -0.12 & 0.03 & 98.13 & -4.88 & $<$0.001 & $<$0.001 & *** & 0.20 & [0.09,1] \\
$\alpha_2$ & N-Back & Stroop & 0.03 & 0.03 & 111.40 & 1.01 & 0.743 & 0.891 & ns & 0.01 & [0.00,1] \\
$\alpha_2$ & Rotation & Stroop & 0.15 & 0.03 & 109.32 & 5.63 & $<$0.001 & $<$0.001 & *** & 0.22 & [0.12,1] \\
\midrule
$\beta_1$ & Chess & N-Back & 0.12 & 0.04 & 101.32 & 2.90 & 0.023 & 0.049 & * & 0.08 & [0.01,1] \\
$\beta_1$ & Chess & Rotation & 0.08 & 0.04 & 99.96 & 1.94 & 0.217 & 0.304 & ns & 0.04 & [0.00,1] \\
$\beta_1$ & Chess & Stroop & 0.23 & 0.04 & 107.61 & 5.41 & $<$0.001 & $<$0.001 & *** & 0.21 & [0.11,1] \\
$\beta_1$ & N-Back & Rotation & -0.04 & 0.04 & 99.80 & -0.89 & 0.809 & 0.928 & ns & 0.01 & [0.00,1] \\
$\beta_1$ & N-Back & Stroop & 0.12 & 0.04 & 107.44 & 2.73 & 0.037 & 0.070 & ns & 0.06 & [0.01,1] \\
$\beta_1$ & Rotation & Stroop & 0.15 & 0.04 & 105.98 & 3.58 & 0.003 & 0.007 & ** & 0.11 & [0.03,1] \\
\midrule
$\beta_2$ & Chess & N-Back & 0.05 & 0.06 & 102.33 & 0.88 & 0.817 & 0.928 & ns & 0.01 & [0.00,1] \\
$\beta_2$ & Chess & Rotation & 0.16 & 0.06 & 101.38 & 2.70 & 0.040 & 0.073 & ns & 0.07 & [0.01,1] \\
$\beta_2$ & Chess & Stroop & 0.26 & 0.06 & 106.56 & 4.32 & $<$0.001 & 0.001 & ** & 0.15 & [0.06,1] \\
$\beta_2$ & N-Back & Rotation & 0.11 & 0.06 & 100.58 & 1.86 & 0.252 & 0.342 & ns & 0.03 & [0.00,1] \\
$\beta_2$ & N-Back & Stroop & 0.21 & 0.06 & 105.76 & 3.53 & 0.003 & 0.008 & ** & 0.11 & [0.03,1] \\
$\beta_2$ & Rotation & Stroop & 0.11 & 0.06 & 104.62 & 1.74 & 0.308 & 0.404 & ns & 0.03 & [0.00,1] \\
\midrule
$\gamma_1$ & Chess & N-Back & 0.13 & 0.05 & 102.04 & 2.47 & 0.071 & 0.111 & ns & 0.06 & [0.01,1] \\
$\gamma_1$ & Chess & Rotation & 0.14 & 0.05 & 100.95 & 2.62 & 0.049 & 0.083 & ns & 0.06 & [0.01,1] \\
$\gamma_1$ & Chess & Stroop & 0.30 & 0.06 & 106.91 & 5.23 & $<$0.001 & $<$0.001 & *** & 0.20 & [0.10,1] \\
$\gamma_1$ & N-Back & Rotation & 0.01 & 0.05 & 100.34 & 0.21 & 0.997 & 0.999 & ns & 0.00 & [0.00,1] \\
$\gamma_1$ & N-Back & Stroop & 0.17 & 0.06 & 106.28 & 2.95 & 0.020 & 0.044 & * & 0.08 & [0.01,1] \\
$\gamma_1$ & Rotation & Stroop & 0.15 & 0.06 & 105.03 & 2.75 & 0.035 & 0.070 & ns & 0.07 & [0.01,1] \\
\midrule
$\gamma_2$ & Chess & N-Back & 0.13 & 0.06 & 102.21 & 2.20 & 0.130 & 0.195 & ns & 0.05 & [0.00,1] \\
$\gamma_2$ & Chess & Rotation & 0.12 & 0.06 & 101.12 & 1.97 & 0.205 & 0.297 & ns & 0.04 & [0.00,1] \\
$\gamma_2$ & Chess & Stroop & 0.28 & 0.06 & 107.11 & 4.52 & $<$0.001 & $<$0.001 & *** & 0.16 & [0.07,1] \\
$\gamma_2$ & N-Back & Rotation & -0.01 & 0.06 & 100.35 & -0.17 & 0.998 & 0.999 & ns & 0.00 & [0.00,1] \\
$\gamma_2$ & N-Back & Stroop & 0.15 & 0.06 & 106.33 & 2.48 & 0.069 & 0.111 & ns & 0.05 & [0.01,1] \\
$\gamma_2$ & Rotation & Stroop & 0.16 & 0.06 & 105.05 & 2.65 & 0.046 & 0.080 & ns & 0.06 & [0.01,1] \\
\bottomrule
\end{tabular}
\end{adjustbox}
\end{table}

\begin{table}[h!]
\caption{Post-hoc contrast results modeling changes in EEG band data as a function of within-task workload for the Chess task. Increases in power were found in among the most separable workload levels (0-3 and 0-2) for all the frequency bands. Increases in power between workload levels 1-3 were likewise found in all except $\beta_1$.}
\label{tab:WITHIN_EEG_PHOCS_CHESS}
\begin{adjustbox}{width=\linewidth}
\begin{tabular}{lccccccccccc}
\toprule
Task & Band & Contrast & Est. & SE & df & t & p & p.adj & p.adj.sig & $\eta_{p}^{2}$ & 95\% CI \\
\midrule
Chess & $\beta_1$ & 0 - 3 & -0.45 & 0.11 & 197.61 & -3.96 & 0.001 & 0.002 & ** & 0.07 & [0.03,1] \\
Chess & $\beta_1$ & 0 - 2 & -0.33 & 0.11 & 197.01 & -3.01 & 0.016 & 0.038 & * & 0.04 & [0.01,1] \\
Chess & $\beta_1$ & 1 - 3 & -0.31 & 0.11 & 197.61 & -2.72 & 0.035 & 0.070 & ns & 0.04 & [0.01,1] \\
Chess & $\beta_1$ & 0 - 1 & -0.14 & 0.11 & 197.01 & -1.27 & 0.586 & 0.669 & ns & 0.01 & [0.00,1] \\
Chess & $\beta_1$ & 1 - 2 & -0.19 & 0.11 & 197.01 & -1.74 & 0.305 & 0.386 & ns & 0.02 & [0.00,1] \\
Chess & $\beta_1$ & 2 - 3 & -0.12 & 0.11 & 197.61 & -1.02 & 0.740 & 0.772 & ns & 0.01 & [0.00,1] \\
\midrule
Chess & $\beta_2$ & 0 - 3 & -0.72 & 0.12 & 197.78 & -6.01 & $<$0.001 & $<$0.001 & *** & 0.15 & [0.08,1] \\
Chess & $\beta_2$ & 0 - 2 & -0.44 & 0.12 & 197.01 & -3.79 & 0.001 & 0.004 & ** & 0.07 & [0.02,1] \\
Chess & $\beta_2$ & 1 - 3 & -0.51 & 0.12 & 197.78 & -4.26 & $<$0.001 & 0.001 & ** & 0.08 & [0.03,1] \\
Chess & $\beta_2$ & 0 - 1 & -0.21 & 0.12 & 197.01 & -1.79 & 0.282 & 0.376 & ns & 0.02 & [0.00,1] \\
Chess & $\beta_2$ & 1 - 2 & -0.23 & 0.12 & 197.01 & -2.00 & 0.193 & 0.289 & ns & 0.02 & [0.00,1] \\
Chess & $\beta_2$ & 2 - 3 & -0.27 & 0.12 & 197.78 & -2.30 & 0.102 & 0.187 & ns & 0.03 & [0.00,1] \\
\midrule
Chess & $\gamma_1$ & 0 - 3 & -0.63 & 0.12 & 197.80 & -5.12 & $<$0.001 & $<$0.001 & *** & 0.12 & [0.06,1] \\
Chess & $\gamma_1$ & 0 - 2 & -0.39 & 0.12 & 197.01 & -3.26 & 0.007 & 0.019 & * & 0.05 & [0.01,1] \\
Chess & $\gamma_1$ & 1 - 3 & -0.49 & 0.12 & 197.80 & -4.00 & 0.001 & 0.002 & ** & 0.07 & [0.03,1] \\
Chess & $\gamma_1$ & 0 - 1 & -0.14 & 0.12 & 197.01 & -1.14 & 0.662 & 0.722 & ns & 0.01 & [0.00,1] \\
Chess & $\gamma_1$ & 1 - 2 & -0.25 & 0.12 & 197.01 & -2.12 & 0.151 & 0.242 & ns & 0.02 & [0.00,1] \\
Chess & $\gamma_1$ & 2 - 3 & -0.23 & 0.12 & 197.80 & -1.92 & 0.223 & 0.315 & ns & 0.02 & [0.00,1] \\
\midrule
Chess & $\gamma_2$ & 0 - 3 & -0.56 & 0.12 & 197.82 & -4.49 & $<$0.001 & 0.001 & *** & 0.09 & [0.04,1] \\
Chess & $\gamma_2$ & 0 - 2 & -0.36 & 0.12 & 197.01 & -2.95 & 0.018 & 0.040 & * & 0.04 & [0.01,1] \\
Chess & $\gamma_2$ & 1 - 3 & -0.46 & 0.12 & 197.82 & -3.71 & 0.002 & 0.005 & ** & 0.07 & [0.02,1] \\
Chess & $\gamma_2$ & 0 - 1 & -0.10 & 0.12 & 197.01 & -0.80 & 0.856 & 0.856 & ns & 0.00 & [0.00,1] \\
Chess & $\gamma_2$ & 1 - 2 & -0.26 & 0.12 & 197.01 & -2.16 & 0.140 & 0.239 & ns & 0.02 & [0.00,1] \\
Chess & $\gamma_2$ & 2 - 3 & -0.20 & 0.12 & 197.82 & -1.60 & 0.382 & 0.459 & ns & 0.01 & [0.00,1] \\
\bottomrule
\end{tabular}
\end{adjustbox}
\end{table}

\begin{table}[h!]
\caption{Post-hoc contrast results modeling EEG band data as a function of within-task workload for the N-Back task. Increases in power were found between workload levels 0-3 for all three bands, and between workload levels and 0-2 for both the $\theta$ and $\alpha_1$ bands.}
\label{tab:WITHIN_EEG_PHOCS_NBACK}
\begin{adjustbox}{width=\linewidth}
\begin{tabular}{lccccccccccc}
\toprule
Task & Band & Contrast & Est. & SE & df & t & p & p.adj & p.adj.sig & $\eta_{p}^{2}$ & 95\% CI \\
\midrule

N-Back & $\theta$ & 0 - 3 & -0.69 & 0.22 & 112.79 & -3.09 & 0.013 & 0.048 & * & 0.08 & [0.02,1] \\
N-Back & $\theta$ & 0 - 2 & -0.70 & 0.22 & 112.63 & -3.20 & 0.010 & 0.048 & * & 0.08 & [0.02,1] \\
N-Back & $\theta$ & 1 - 3 & -0.45 & 0.23 & 112.24 & -1.97 & 0.207 & 0.339 & ns & 0.03 & [0.00,1] \\
N-Back & $\theta$ & 0 - 1 & -0.24 & 0.22 & 113.63 & -1.08 & 0.703 & 0.953 & ns & 0.01 & [0.00,1] \\
N-Back & $\theta$ & 1 - 2 & -0.46 & 0.22 & 115.15 & -2.05 & 0.177 & 0.319 & ns & 0.04 & [0.00,1] \\
N-Back & $\theta$ & 2 - 3 & 0.01 & 0.22 & 114.33 & 0.06 & 1.000 & 1.000 & ns & 0.00 & [0.00,1] \\
\midrule
N-Back & $\alpha_1$ & 0 - 3 & -0.67 & 0.21 & 111.28 & -3.22 & 0.009 & 0.048 & * & 0.09 & [0.02,1] \\
N-Back & $\alpha_1$ & 0 - 2 & -0.70 & 0.20 & 111.20 & -3.42 & 0.005 & 0.048 & * & 0.10 & [0.03,1] \\
N-Back & $\alpha_1$ & 1 - 3 & -0.58 & 0.21 & 110.28 & -2.76 & 0.034 & 0.077 & ns & 0.06 & [0.01,1] \\
N-Back & $\alpha_1$ & 0 - 1 & -0.08 & 0.21 & 111.94 & -0.41 & 0.977 & 1.000 & ns & 0.00 & [0.00,1] \\
N-Back & $\alpha_1$ & 1 - 2 & -0.61 & 0.21 & 113.12 & -2.93 & 0.021 & 0.063 & ns & 0.07 & [0.01,1] \\
N-Back & $\alpha_1$ & 2 - 3 & 0.03 & 0.21 & 112.48 & 0.15 & 0.999 & 1.000 & ns & 0.00 & [0.00,1] \\
\midrule
N-Back & $\alpha_2$ & 0 - 3 & -0.70 & 0.22 & 112.79 & -3.14 & 0.012 & 0.048 & * & 0.08 & [0.02,1] \\
N-Back & $\alpha_2$ & 0 - 2 & -0.47 & 0.22 & 112.64 & -2.15 & 0.143 & 0.286 & ns & 0.04 & [0.00,1] \\
N-Back & $\alpha_2$ & 1 - 3 & -0.63 & 0.23 & 112.25 & -2.75 & 0.034 & 0.077 & ns & 0.06 & [0.01,1] \\
N-Back & $\alpha_2$ & 0 - 1 & -0.07 & 0.23 & 113.64 & -0.33 & 0.988 & 1.000 & ns & 0.00 & [0.00,1] \\
N-Back & $\alpha_2$ & 1 - 2 & -0.40 & 0.23 & 115.16 & -1.77 & 0.291 & 0.436 & ns & 0.03 & [0.00,1] \\
N-Back & $\alpha_2$ & 2 - 3 & -0.23 & 0.22 & 114.33 & -1.01 & 0.741 & 0.953 & ns & 0.01 & [0.00,1] \\
\bottomrule
\end{tabular}
\end{adjustbox}
\end{table}